\documentclass[sigconf, 9pt]{acmart}

\usepackage{algorithm}
\usepackage{algpseudocode}

\let\oldReturn\Return
\renewcommand{\Return}{\State\oldReturn}
\usepackage{listings}

\AtBeginDocument{%
  \providecommand\BibTeX{{%
    \normalfont B\kern-0.5em{\scshape i\kern-0.25em b}\kern-0.8em\TeX}}}

\copyrightyear{2022}
\acmYear{2022}
\setcopyright{acmcopyright}\acmConference[ICS '22]{2022 International Conference
on Supercomputing}{June 28--30, 2022}{Virtual Event, USA}
\acmBooktitle{2022 International Conference on Supercomputing (ICS '22), June 28--30, 2022, Virtual Event, USA}
\acmPrice{15.00}
\acmDOI{10.1145/3524059.3532363}
\acmISBN{978-1-4503-9281-5/22/06}

\usepackage{comment}
\usepackage[inline]{enumitem}
\usepackage{multirow}
\usepackage{caption}
\usepackage{subcaption}
\usepackage{xcolor,colortbl}
\usepackage{booktabs, makecell}
\usepackage{threeparttable}

\usepackage{graphicx}
\usepackage{pgfplots}
\pgfplotsset{compat=1.16}
\usepackage{tikz}
\usetikzlibrary{patterns}
\usetikzlibrary{shapes}
\usepackage[nolist]{acronym}
\usepackage{balance}

\newcommand{\SCAL}[1]{$#1$}
\newcommand{\MATRIX}[1]{$\textbf{#1}$}
\newcommand{\MATELEM}[3]{${#1}_{#2,#3}$}
\newcommand{\VECTOR}[1]{$\textbf{#1}$}
\newcommand{\VECELEM}[2]{${#1}_{#2}$}
\newcommand{\TENSOR}[1]{$\mathcal{#1}$}

\newcommand{\TENELEM}[3]{$#1_{#2_{1},\dots,#2_{#3}}$}
\newcommand{\MATTEN}[2]{$\textbf{#1}_{(#2)}$}

\newcommand{\REALONE}[1]{$\mathbb{R}^{#1}$}
\newcommand{\REALTWO}[2]{$\mathbb{R}^{#1\times #2}$}
\newcommand{\REALTHREE}[3]{$\mathbb{R}^{#1\times #2 \times #3}$}
\newcommand{\REALX}[2]{$\mathbb{R}^{#1_{1}\times \cdots \times #1_{#2}}$}

\newcommand{\IGNORE}[1]{}

\definecolor{greyback}{RGB}{248,248,248}
\definecolor{deepblue}{RGB}{43,131,186}
\definecolor{greenalt}{RGB}{35,139,69}
\definecolor{darkred}{RGB}{215,25,28}
\definecolor{darkorange}{RGB}{253,174,97}

\definecolor{Gray}{gray}{0.85}
\newcolumntype{a}{>{\columncolor{Gray}}c}

\begin{document}

\title{Efficient, Out-of-Memory Sparse MTTKRP on Massively Parallel Architectures}

\author{Andy Nguyen}
\affiliation{%
  \institution{University of Oregon}\country{}}
\email{andyn@uoregon.edu}

\author{ Ahmed E. Helal}
\affiliation{%
  \institution{Intel Labs}\country{}}
\email{ahmed.helal@intel.com}

\author{Fabio Checconi}
\affiliation{%
  \institution{Intel Labs}\country{}}
\email{fabio.checconi@intel.com}

\author{Jan Laukemann}
\affiliation{%
  \institution{University of Erlangen-Nürnberg}\country{}}
\email{jan.laukemann@fau.de}
\authornote{The author was employed by Intel when this research was conducted.}

\author{Jesmin Jahan Tithi}
\affiliation{%
  \institution{Intel Labs}\country{}}
\email{jesmin.jahan.tithi@intel.com}

\author{Yongseok Soh}
\affiliation{%
  \institution{University of Oregon}\country{}}
\email{ysoh@uoregon.edu}

\author{Teresa Ranadive}
\affiliation{%
 \institution{Laboratory for Physical Sciences}\country{}}
\email{tranadive@lps.umd.edu}

\author{Fabrizio Petrini}
\affiliation{%
  \institution{Intel Labs}\country{}}
\email{fabrizio.petrini@intel.com}

\author{Jee W. Choi}
\affiliation{%
  \institution{University of Oregon}\country{}}
\email{jeec@uoregon.edu}
\renewcommand{\shortauthors}{Andy Nguyen, Ahmed E. Helal, Fabio Checconi, et al.}

\begin{abstract}
Tensor decomposition (TD) is an important method for extracting latent information from high-dimensional (multi-modal) \emph{sparse} data. 
This study presents a novel framework for accelerating fundamental TD operations on massively parallel GPU architectures.
In contrast to prior work, the proposed Blocked Linearized CoOrdinate (BLCO) format enables efficient \emph{out-of-memory} computation of tensor algorithms using a \emph{unified implementation} that works on a \emph{single tensor copy}.
Our adaptive blocking and linearization strategies not only meet the resource constraints of GPU devices, but also accelerate data indexing, eliminate control-flow and memory-access irregularities, and reduce kernel launching overhead.
To address the substantial synchronization cost on GPUs, we introduce an opportunistic conflict resolution algorithm, in which threads collaborate instead of contending on memory access to discover and resolve their conflicting updates \emph{on-the-fly}, without keeping any auxiliary information or storing non-zero elements in specific mode orientations. 
As a result, our framework delivers superior in-memory performance compared to prior state-of-the-art, and is the \emph{only} framework capable of processing out-of-memory tensors.
On the latest Intel and NVIDIA GPUs, BLCO achieves $2.12-2.6\times$ geometric-mean speedup (with up to $33.35\times$ speedup) over the state-of-the-art mixed-mode compressed sparse fiber (MM-CSF) on a range of real-world sparse tensors.
\end{abstract}



\maketitle

\section{Introduction}
\label{sec:introduction}
Tensors are higher-order generalization of matrices, and they provide a natural abstraction for complex and inter-related data.
Many critical applications, such as data mining~\cite{Kolda2008,Papalexakis2016}, social network analytics~\cite{Rettinger2012,Fernandes2019}, cybersecurity~\cite{Fanaee-T2016,Bruns-Smith2016}, and healthcare~\cite{Ho2014,He2019}, generate massive amounts of multi-dimensional (multi-modal) data as \emph{sparse tensors} that can be analyzed quickly and efficiently using \emph{tensor decomposition} (TD).
The most popular TD method is the canonical polyadic decomposition (CPD) model, which approximates a tensor as a sum of a finite number of rank-one tensors such that each rank-one tensor corresponds to a useful data property~\cite{Kolda2009,Bader2007}.
Computing the CPD of a sparse tensor is typically dominated by the \emph{matricized tensor times Khatri-Rao product} (MTTKRP) operation, which makes up approximately $90\%$ of the total execution time~\cite{Smith2015}.

TD algorithms for high-dimensional sparse data are challenging to execute on emerging parallel architectures due to their low arithmetic intensity, irregular memory access, workload imbalance, and synchronization overhead~\cite{Choi2018, ahmed2021}. 
To improve the performance of these memory-bound workloads, recent studies~\cite{Liu2017,Nisa2019a,Nisa2019b,Dun2021,phipps2019software} exploit massively parallel architectures equipped with High Bandwidth Memory (HBM), namely GPUs, to accelerate the MTTKRP kernel. 
While such accelerators deliver memory bandwidth exceeding $2$ TB/s~\cite{nvidia2021}, they suffer from limited memory capacity and high memory-access latency, which is in the order of hundreds of processor cycles~\cite{mei2016dissecting, jia2018dissecting}. 
Moreover, the high memory latency together with the massive number of threads can substantially increase the synchronization overhead.

Therefore, the prior studies focus primarily on designing \emph{sparse formats} that \emph{compress} the tensor to decrease its memory footprint and/or \emph{group} data-dependent non-zero elements to \emph{reduce atomic operations}, which are particularly more expensive on massively parallel architectures. 
However, these strategies result in formats that are \emph{mode-specific}, where non-zero elements are organized/accessed according to a specific mode (i.e., dimension) orientation. 
Mode-specific formats typically require keeping multiple tensor copies~\cite{Liu2017,Nisa2019a, Dun2021} and/or extra mapping and scheduling information (e.g., flags or arrays) about groups of data-dependent non-zero elements for \emph{every} mode~\cite{Liu2017, phipps2019software, Dun2021}, which can significantly increase their overall memory footprint. 

\begin{figure}[tpb]
\centering
\includegraphics[width=1.0\linewidth]{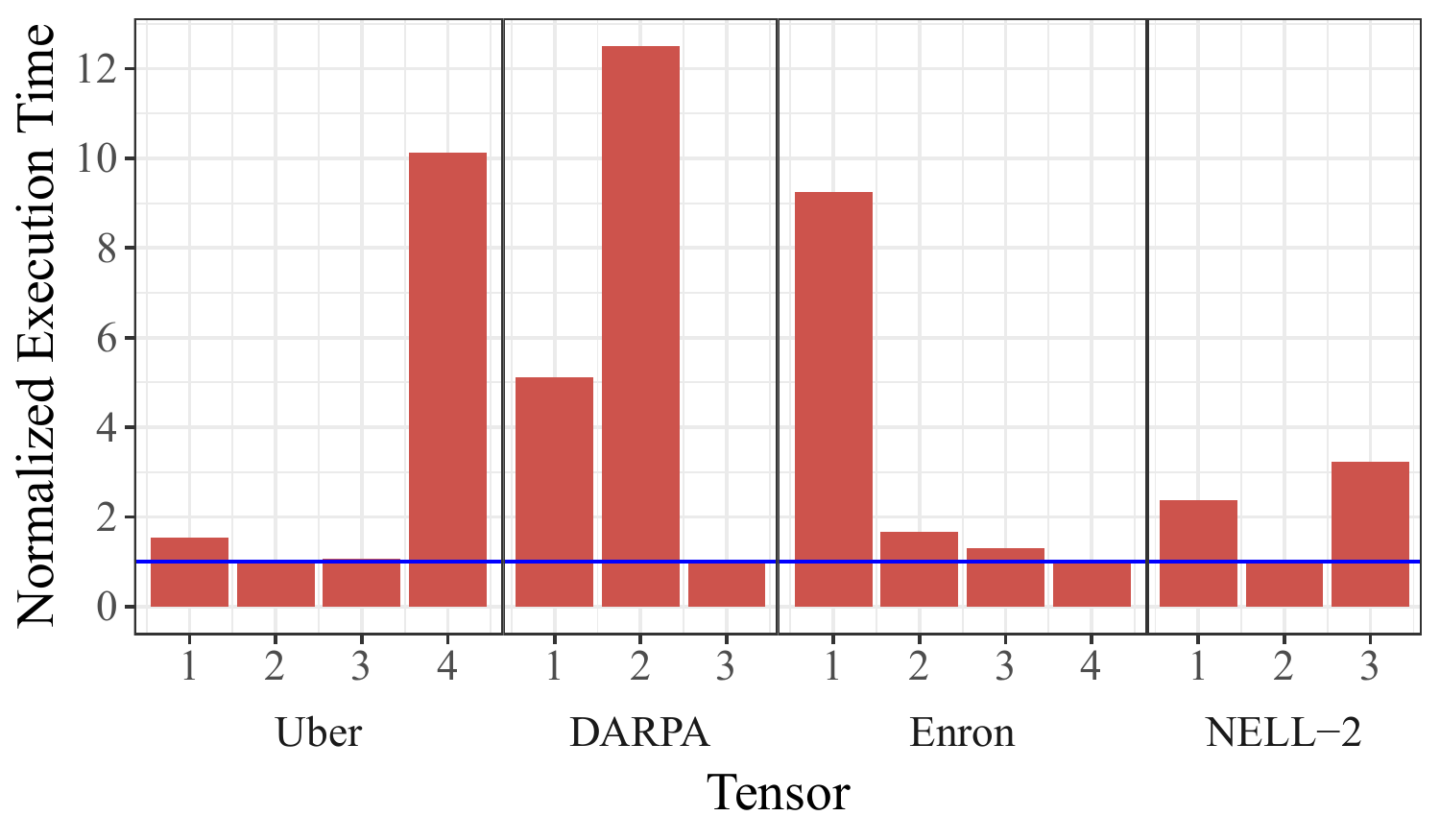}
	\vspace*{-25pt}
	\caption{The MTTKRP execution time of MM-CSF\protect\footnotemark across all modes on the A100 GPU, normalized by the lowest execution time (denoted by the blue line) for each data set.
	The decomposition rank is $32$.
	For NELL-2, modes 1 and 3 take $2$--$3\times$ longer to execute than mode 2, 
	while for Uber and Enron, one mode takes \emph{more than $9\times$ longer} to execute.
	For DARPA, mode $1$ and mode $2$ take $5\times$ and $12\times$ longer, respectively, than mode $3$.
	Note that the number of FLOPs computed is \emph{identical} across modes for each data set.
    	}
  \vspace*{-5pt}
\label{fig:motivation}
\end{figure}

\footnotetext{https://github.com/isratnisa/MM-CSF}
Furthermore, in such mode-specific tensor formats, high compression along one mode may result in poor performance along other modes~\cite{Nisa2019b}.
For example, Figure~\ref{fig:motivation} illustrates the impact of mode-specific compression on the performance of MTTKRP across all modes for the state-of-the-art mixed-mode compressed sparse fiber (MM-CSF) format~\cite{Nisa2019b}, which currently has the best performance on GPUs, achieving $2-1.4\times$ average speedup over prior GPU formats.
This result demonstrates that depending on the data set, the execution time may vary by an \emph{order of magnitude} across modes when the compression favors one particular mode over others.

Additionally, data formats based on compressed sparse fiber (CSF), e.g., balanced CSF (B-CSF)~\cite{Nisa2019a} and MM-CSF~\cite{Nisa2019b}, require mode-specific implementations for tensor operations, such as MTTKRP, as their tree-like data structure necessitates different methods of tree traversal and intermediate result accumulation for each mode.
This leads to poor code scalability and portability.
Lastly, the mode-specific nature of these formats (i.e., tree-based structure and/or auxiliary copies/scheduling data) makes out-of-memory (OOM) tensors (i.e., large-scale tensors that do not fit in GPU memory) difficult to process.
As a result, current GPU frameworks for MTTKRP are constrained to sparse tensors that can fit in the limited device memory (i.e., in-memory) and lack support for real-world OOM tensors with billions of non-zero elements.
 
To summarize, the state-of-the-art tensor decomposition approaches for massively parallel GPU architectures 
rely on \emph{mode-specific} formats to reduce data movement via \emph{compression} and to decrease the number of atomic operations by \emph{reordering} of non-zero elements.
However, these techniques can result in
\begin{enumerate*}[label=(\roman*)]
    \item drastic performance loss and variations along different tensor modes, 
    \item significant memory overhead to keep extra tensor copies or mapping flags/arrays,
    \item complex algorithms and code implementations to handle tensor operations across different mode orientations, and
    \item limited support for real-world data sets due to inability to process OOM tensors on memory-constrained accelerators
\end{enumerate*}.

To address these limitations, 
we propose a novel \emph{mode-agnostic} framework for large-scale sparse tensor decomposition on GPUs.
We make four key contributions:
\begin{itemize}
    \item We analyze prior state-of-the-art sparse tensor formats and parallel MTTKRP algorithms to determine the key performance bottlenecks and limitations (Section~\ref{sec:challenges}).
    \item We introduce Blocked Linearized CoOrdinate (BLCO), a new sparse tensor format that uses an adaptive blocking and linearization approach to generate coarse-grained tensor blocks,
    based on the resource constraints of target devices, while exposing the fine-grained parallelism within a block to efficiently utilize the accelerator hardware. 
    Our BLCO format accelerates indexing, reduces data movement, decreases kernel launch overhead, supports out-of-memory tensors, and enables a unified tensor representation and MTTKRP implementation (Section~\ref{sec:blco}). 
    \item We present a novel massively parallel MTTKRP algorithm 
    that eliminates irregularities in control-flow and memory-access, while efficiently discovering and resolving conflicting updates across threads \emph{on-the-fly}.
    By employing cooperative thread teams and using low-latency registers/memories, our algorithm reduces synchronization cost \emph{without} requiring any mode-specific information or storing the data in a particular mode order 
    (Section~\ref{sec:hreduction}).
    \item We demonstrate substantial performance improvement compared to prior state of the art, achieving $2.12-2.6\times$ geometric-mean speedup (up to $33.35\times$ speedup) across the latest Intel and NVIDIA GPUs
    for a representative set of real-world tensors.
    Furthermore, we show the utility of BLCO in processing out-of-memory tensors in contrast to existing GPU-based TD frameworks (Section~\ref{sec:exp}).
\end{itemize}

\section{Background}
\label{sec:background}
In this section, we provide a brief overview of tensors, their decomposition, and related notations.
For more details on tensor decomposition, we direct the reader towards the work by Kolda and Bader~\cite{Kolda2009,Bader2007}.  
 
\subsection{Notation}
\label{subsec:notations}
Tensors are multi-modal arrays that generalize the concepts of vectors and matrices.
An $N$-order tensor is an array with $N$ modes.
We use the following notation in this paper:
\begin{enumerate}
	\item
		\emph{Scalars} are written with lowercase letters  (e.g., \SCAL{a}).
	\item
		\emph{Vectors} (first-order tensors) are written with bold lowercase letters (e.g., \VECTOR{a}~$\in$~\REALONE{I}).
		The $i^{th}$ entry of \VECTOR{a}~$\in$~\REALONE{I} is denoted \VECELEM{a}{i}.
	\item
		\emph{Matrices} (second-order tensors) are written with bold capital letters (e.g., \MATRIX{A}~$\in$~\REALTWO{I}{J}). 
		The $(i,j)^{th}$ entry of \MATRIX{A}~$\in$~\REALTWO{I}{J} is denoted \MATELEM{a}{i}{j}.
	\item
		\emph{Higher-order tensors} are written with Euler script letters (e.g., \TENSOR{X} $\in$ \REALX{I}{N}).  The $(i_{1},\dots,i_{N})^{th}$ entry of the $N$-order tensor \TENSOR{X} $\in$ \REALX{I}{N} is denoted \TENELEM{x}{i}{N}.
	\item
		\emph{Fibers} are the analogue of matrix rows/columns for higher-order tensors.
		A mode-$n$ fiber of a tensor \TENSOR{X} is any vector formed by fixing all indices of \TENSOR{X}, \emph{except} the $n^{th}$ index
		(e.g., a matrix column is defined by fixing the second index, and is therefore a mode-$1$ fiber).
	\item
	    \emph{Hadamard} product is an element-wise product between two vectors or matrices, and is denoted by the symbol ``$*$''.
	\item 
	    \emph{Kronecker} product between two matrices \MATRIX{A} $\in$ \REALTWO{I}{J} and \MATRIX{B} $\in$ \REALTWO{K}{L} produces the matrix \MATRIX{C} $\in$ \REALTWO{IJ}{KL}, where 
	    \begin{equation*}
	        \textbf{C} = 
	        \begin{bmatrix} 
	        \SCAL{a}_{1,1}\MATRIX{B} & \SCAL{a}_{1,2}\MATRIX{B} & \cdots & \SCAL{a}_{1,J}\MATRIX{B}\\
	        \SCAL{a}_{2,1}\MATRIX{B} & \SCAL{a}_{2,2}\MATRIX{B} & \cdots & \SCAL{a}_{2,J}\MATRIX{B}\\
	        \vdots & \vdots & \ddots & \vdots\\
	        \SCAL{a}_{I,1}\MATRIX{B} & \SCAL{a}_{I,2}\MATRIX{B} & \cdots & \SCAL{a}_{I,J}\MATRIX{B}\\
	        \end{bmatrix}
	    \end{equation*}
	    and is denoted \MATRIX{A} $\otimes$ \MATRIX{B}.
\end{enumerate}

\subsection{Canonical Polyadic Decomposition}
\label{subsec:cp-als}
\begin{figure}[htb!]
  \centering
  \includegraphics[width=1\linewidth]{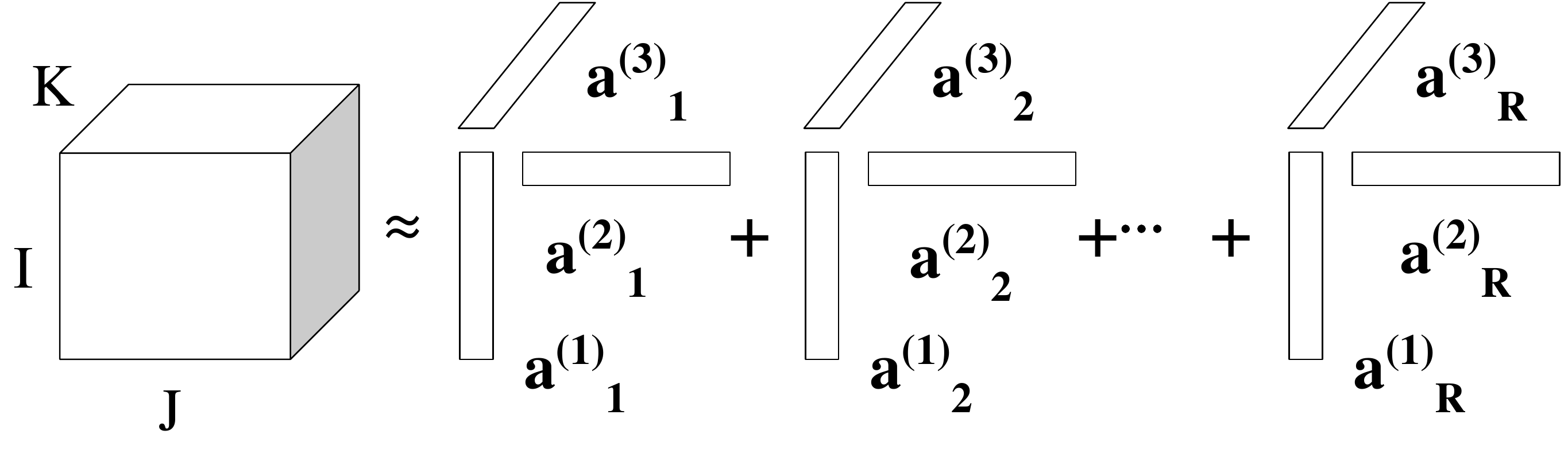}
     \vspace*{-22pt}
  \caption{Rank-$R$ CPD of a third-order tensor. The factor matrix \MATRIX{A}$^{(1)}$ consists of
  vectors \VECTOR{a}$^{(1)}_{1}$, \VECTOR{a}$^{(1)}_{2}$, $\cdots$, \VECTOR{a}$^{(1)}_{R}$.
  }
  \label{fig:cpd}
\end{figure}

The canonical polyadic decomposition (CPD) is a widely used tensor factorization model in data analysis.
CPD approximates an $N$-order tensor \TENSOR{X} by the sum of $R$ outer products of $N$ appropriately sized vectors, for some {\it a priori} chosen $R$. 
Each of the $N$ vectors in a given outer product corresponds to a particular tensor mode.  The outer products are called the rank-$1$ tensors of the decomposition and the quantity $R$ is called the decomposition \emph{rank}.   
By arranging the $R$ vectors corresponding to a particular mode as the columns of a matrix,  we obtain the \emph{factor matrix} associated with that mode. The decomposition of \TENSOR{X} can then be written in terms of its factor matrices.
For example, if \TENSOR{X}~$\in$~\REALTHREE{I}{J}{K} is a third-order tensor, the CPD of \TENSOR{X} may be written in terms of three factor matrices \MATRIX{A}$^{(1)}$~$\in$~\REALTWO{I}{R}, \MATRIX{A}$^{(2)}$~$\in$~\REALTWO{J}{R}, and \MATRIX{A}$^{(3)}$~$\in$~\REALTWO{K}{R}
, as shown in Figure~\ref{fig:cpd}.

CANDECOMP/PARAFAC alternating least squares~(CP-ALS) algorithm is a popular method for calculating CPD.
During each CP-ALS iteration, we update the factor matrix corresponding to a given tensor mode by solving a linear least squares problem, while fixing the remaining factor matrices; this is done exactly once for each factor matrix per iteration.
The most expensive operation of CP-ALS, as well as many other tensor algorithms, is the  the matricized tensor times Khatri-Rao product (MTTKRP)~\cite{Smith2015}. 
The CP-ALS algorithm for an $N$-order tensor is detailed in Algorithm~\ref{alg:cp-als}.
Line~\ref{line:mttkrp} shows the MTTKRP computations.

\begin{algorithm}[htb]
\begin{algorithmic}[1]
\Require An $N$-order sparse tensor \TENSOR{X} $\in$ \REALX{I}{N}, randomly initialized dense factor matrices \MATRIX{A}$^{(1)}$  $\in$ \REALTWO{I_{1}}{R}, \MATRIX{A}$^{(2)}$  $\in$ \REALTWO{I_{2}}{R}, $\cdots$, \MATRIX{A}$^{(N)}$  $\in$ \REALTWO{I_{N}}{R}.
\Ensure Updated factor matrices that approximate \TENSOR{X}.
\Repeat
    \For{ $n = 1, \dots, N$}
    \State \MATRIX{V} $\gets$ \MATRIX{A}$^{(1)T}$\MATRIX{A}$^{(1)}$ $*$ $\cdots$ $*$ \MATRIX{A}$^{(n-1)T}$\MATRIX{A}$^{(n-1)}$ $*$ \par\hspace{\algorithmicindent}\hspace{\algorithmicindent} \MATRIX{A}$^{(n+1)T}$\MATRIX{A}$^{(n+1)}$ $*$ $\cdots$ $*$  \MATRIX{A}$^{(N)T}$\MATRIX{A}$^{(N)}$\par
    \State \MATRIX{M} $\gets$ $\textbf{X}_{(n)} (\textbf{A}^{(N)} \odot  \cdots\odot \textbf{A}^{(n+1)} \odot \textbf{A}^{(n-1)} \odot  \cdots \odot \textbf{A}^{(1)})$\label{line:mttkrp}
    \State \MATRIX{A}$^{(n)}$ $\gets$ \MATRIX{M} \MATRIX{V}$^{\dagger}$ \Comment{$\dagger$ denotes the pseudo-inverse}
    \label{line:a_n_update}
    \EndFor
\Until {fit ceases to improve or maximum \# of iterations reached}
\Return \MATRIX{A}$^{(1)}$, $\cdots$, \MATRIX{A}$^{(N)}$
\end{algorithmic}
\caption{The CP-ALS algorithm.}
\label{alg:cp-als}
\end{algorithm}

\subsection{Sparse MTTKRP}
\label{subsec:mttkrp}
The matricized tensor times Khatri-Rao product (MTTKRP) kernel involves two basic operations: 
\begin{enumerate}
	\item 
	\emph{Tensor matricization} is the process in which a tensor is unfolded into a matrix.
	The mode-$n$ matricization of a tensor \TENSOR{X}, denoted by \MATTEN{X}{n}, is obtained by laying out the mode-$n$ fibers of \TENSOR{X} as the columns of \MATTEN{X}{n}.
	\item
	The \emph{Khatri-Rao product}~\cite{KR_product} is the ``matching column-wise'' Kronecker product of two matrices.
	Given \MATRIX{A} $\in$ \REALTWO{I}{R} and \MATRIX{B} $\in$ \REALTWO{J}{R}, their Khatri-Rao product is 
	\MATRIX{K} $=$ \MATRIX{A} $\odot$ \MATRIX{B},
	where
		\MATRIX{A} $\odot$ \MATRIX{B} = $\left[\textbf{a}_{1}\otimes \textbf{b}_{1}\ \textbf{a}_{2}\otimes \textbf{b}_{2} \dots \textbf{a}_{R}\otimes \textbf{b}_{R}\right] \nonumber$ $\in$ \REALTWO{I\cdot J}{R}.
\end{enumerate}

For an $N$-order tensor \TENSOR{X} and factor matrices \MATRIX{A}$^{(1)}$, \MATRIX{A}$^{(2)}$, $\cdots$, \MATRIX{A}$^{(N)}$, the mode-$n$ MTTKRP is given by
\begin{equation}\label{eqn:MTTKRP}
\textbf{M} = \textbf{X}_{(n)} (\textbf{A}^{(N)} \odot  \cdots\odot \textbf{A}^{(n+1)} \odot \textbf{A}^{(n-1)} \odot  \cdots \odot \textbf{A}^{(1)}).  
\end{equation}

\begin{figure}[htb]
  \centering
  \includegraphics[width=0.8\linewidth]{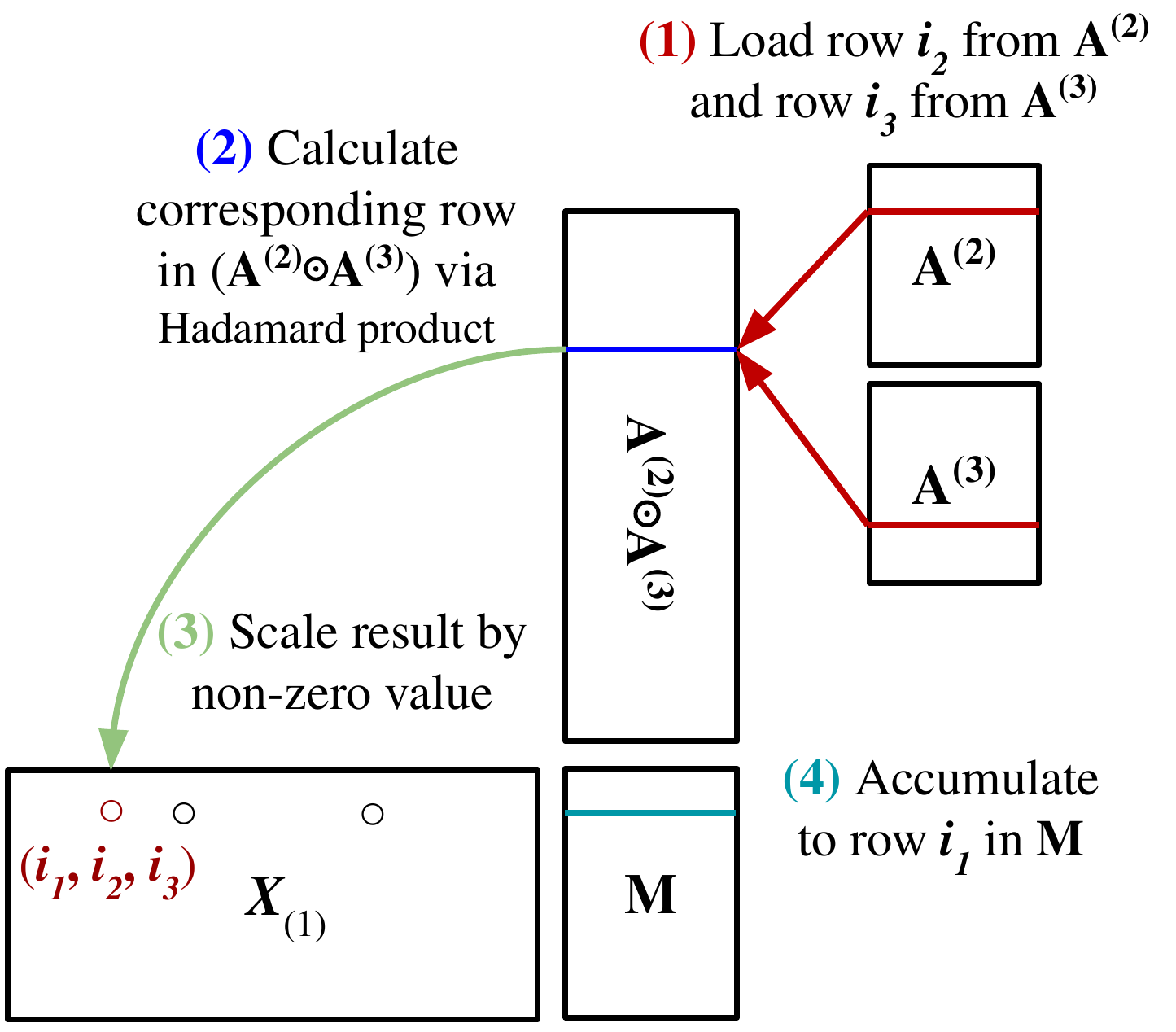}
     \vspace*{-8pt}
  \caption{Mode-$1$ MTTKRP operation for a third-order sparse tensor.
  For each non-zero element with index ($i_{1}$, $i_{2}$, $i_{3}$), rows $i_{2}$ and $i_{3}$ from factor matrices $\textbf{A}^{(2)}$ and $\textbf{A}^{(3)}$ are fetched~(1) and their Hadamard product (element-wise product) is calculated~(2).
  The result is scaled by the non-zero element's value~(3), and then accumulated to matrix row $i_{1}$ from matrix $\textbf{M}$~(4), which is later used to calculate $\textbf{A}^{(1)}$.}
  \label{fig:mttkrp}
   \vspace*{-8pt}
\end{figure}

\MATRIX{M} has the same number of rows and columns as \MATRIX{A}$^{(n)}$ and it is utilized to update \MATRIX{A}$^{(n)}$ in CP-ALS (Line~\ref{line:a_n_update} in Algorithm~\ref{alg:cp-als}).  
Note that for \emph{sparse} tensors, explicitly forming and multiplying \MATTEN{X}{n} and the corresponding Khatri-Rao product (cf.~(\ref{eqn:MTTKRP})) is expensive and unnecessary.
In practice~\cite{Bader2007,Smith2015,li2018hicoo,ahmed2021}, MTTKRP is calculated using the factor matrix rows corresponding to the non-zero elements of the tensor only as needed (c.f.~Figure~\ref{fig:mttkrp}).

\section{Sparse Tensor Formats for GPUs}
\label{sec:challenges}
The state-of-the-art tensor formats for massively parallel GPUs use list-based~\cite{Liu2017,Dun2021,phipps2019software} or tree-based~\cite{Nisa2019a,Nisa2019b} data structures to store high-dimensional sparse data in a \emph{mode-specific} form. In this section, we provide an overview of the flagged coordinate (F-COO) and MM-CSF formats, which are representative of the two main format categories for GPU architectures.
In contrast to sparse linear algebra, higher-order tensor algorithms typically perform tensor operations on \emph{every mode orientation}. Using MTTKRP as a case study, we illustrate the challenges in optimizing and efficiently executing sparse tensor operations on GPUs.

\subsection{F-COO Format}
F-COO is an example of list-based formats that explicitly store non-zero elements with their coordinate indices.
The simplest form of this category is the coordinate (COO) format, which keeps $N$+$1$ lists for an $N$-order tensor---$N$ lists for the indices, and one list for the non-zero values.
While COO is a \emph{mode-agnostic} format, it suffers from substantial synchronization overhead due to update conflicts across threads.

For example, Figure~\ref{fig:mttkrp} depicts the mode-1 MTTKRP operation, where every thread that is processing non-zero elements with mode-1 index of $i_{1}$ (red and black circles) accumulates its partial result to row $i_{1}$ of matrix \MATRIX{M} (step (4)).
This results in chains of read-after-write (RAW) data hazards, which require expensive locks or atomic operations to resolve;
on GPUs, this can quickly become a severe performance bottleneck because of the large number of concurrent threads and the high-latency memory system.

\begin{figure}[htbp]
     \centering
     \begin{subfigure}[b]{0.2\textwidth}
         \centering
         \begin{tabular}{|c|c|c||c|}
            \hline
             $i_{1}$ & $i_{2}$ & $i_{3}$ & v \\
             \hline
             1 & 1 & 1 & 1.0\\
             \hline
             1 & 1 & 2 & 2.0\\
             \hline             
             1 & 3 & 3 & 3.0\\
             \hline
             2 & 1 & 2 & 4.0\\
             \hline             
             2 & 1 & 3 & 5.0\\
             \hline
             3 & 1 & 2 & 6.0\\
             \hline             
             3 & 4 & 4 & 7.0\\
             \hline
             4 & 2 & 1 & 8.0\\
             \hline             
             4 & 2 & 2 & 9.0\\
             \hline             
             4 & 3 & 3 & 10.0\\
             \hline
             4 & 3 & 4 & 11.0\\             
             \hline
             4 & 4 & 4 & 12.0\\             
             \hline
         \end{tabular}
         \caption{COO.}
         \label{fig:coo}
     \end{subfigure}
     \hfill
     \begin{subfigure}[b]{0.25\textwidth}
         \centering
         \begin{tabular}{r|c|c|c||c|}
            \cline{2-5}
             & $bf$ & $i_{2}$ & $i_{3}$ & v \\
             \hline
             \multirow{4}{*}{$sf$=1} & 1 & 1 & 1 & 1.0\\
             \cline{2-5}
             & 1 & 1 & 2 & 2.0\\
             \cline{2-5}             
             & 0 & 3 & 3 & 3.0\\
             \cline{2-5}
             & 1 & 1 & 2 & 4.0\\
             \hline             
             \multirow{4}{*}{$sf$=1} & 0 & 1 & 3 & 5.0\\
             \cline{2-5}
             & 1 & 1 & 2 & 6.0\\
             \cline{2-5}
             & 0 & 4 & 4 & 7.0\\
             \cline{2-5}             
             & 1 & 2 & 1 & 8.0\\
             \hline
             \multirow{4}{*}{$sf$=0} & 1 & 2 & 2 & 9.0\\
             \cline{2-5}             
             & 1 & 3 & 3 & 10.0\\
             \cline{2-5}
             & 1 & 3 & 4 & 11.0\\
             \cline{2-5}             
             & 0 & 4 & 4 & 12.0\\
             \hline
         \end{tabular}
         \caption{F-COO for mode-$1$ MTTKRP.}
         \label{fig:f-coo}
     \end{subfigure}
        \vspace*{-8pt}
        \caption{Comparison between COO~(\ref{fig:coo}) and F-COO~(\ref{fig:f-coo}).
        In F-COO, the target mode index ($i_{1}$) is replaced by $bf$ (bit flag) that goes from $1$ to $0$ when the index changes.
        The $sf$ (start flag) stores a bit for each non-zero group, where $1$ indicates that a new target index has been encountered by the group.
        }
        \label{fig:comparison-f-coo}
          \vspace*{-8pt}
\end{figure}

To address this issue, F-COO~\cite{Liu2017} stores sparse tensors in a sorted \emph{mode-specific} form to \emph{group} non-zero elements with the same $i_{1}$ index together (same $i_{1}$ for mode-$1$ MTTKRP, same $i_{2}$ for mode-$2$ MTTKRP, etc.), so that the accumulation of partial results can be performed locally using segmented scan~\cite{sengupta2008efficient, yan2013streamscan} and globally using atomic operations when group boundaries are crossed.
Additionally, F-COO keeps extra mapping/scheduling information (flags) for synchronization to delineate the end of a non-zero group.
Thus, the F-COO format needs to keep $N$ tensor copies in the GPU memory for an $N$-order tensor. 
As a result, while F-COO reduces the number of global atomic operations, it has high memory usage due to extra data.
Figure~\ref{fig:comparison-f-coo} shows an example sparse tensor in the COO~(\ref{fig:coo}) and F-COO~(\ref{fig:f-coo}) representations.

\subsection{MM-CSF Format}
MM-CSF represents a class of compressed formats that use tree-based structures to encode higher-order tensors.
The original CSF format~\cite{Smith2015, Smith2015a} extends traditional compressed matrix formats, such as the compressed sparse row (CSR) format, by storing a tensor as a collection of index sub-trees with mode-specific ordering.
Given a CSF representation with a mode ordering of $1$-$2$-$3$, where $1$ is the root mode and $3$ is the leaf mode, the root node of each sub-tree represents the factor matrix row that will be updated during \emph{mode-$1$} MTTKRP, and the leaf nodes represent the non-zero elements that contribute to that update.

Thus, to calculate MTTKRP for all modes, multiple CSF copies with different root modes are needed, which increases the memory footprint by a factor of $N$, where $N$ is the number of modes.
Alternatively, the index sub-trees can be traversed both bottom-up and top-down, meeting at the tree level with the target mode for MTTKRP.
While the second approach allows computing MTTKRP for all modes with one copy of the CSF format, it requires expensive synchronization to avoid update conflicts as well as separate tree traversal implementations.
Moreover, regardless of the strategy used, CSF suffers from \emph{workload imbalance} due to the variable size of each sub-tree. 

\begin{figure}[htbp]
    \centering
    \includegraphics[width=0.75\linewidth]{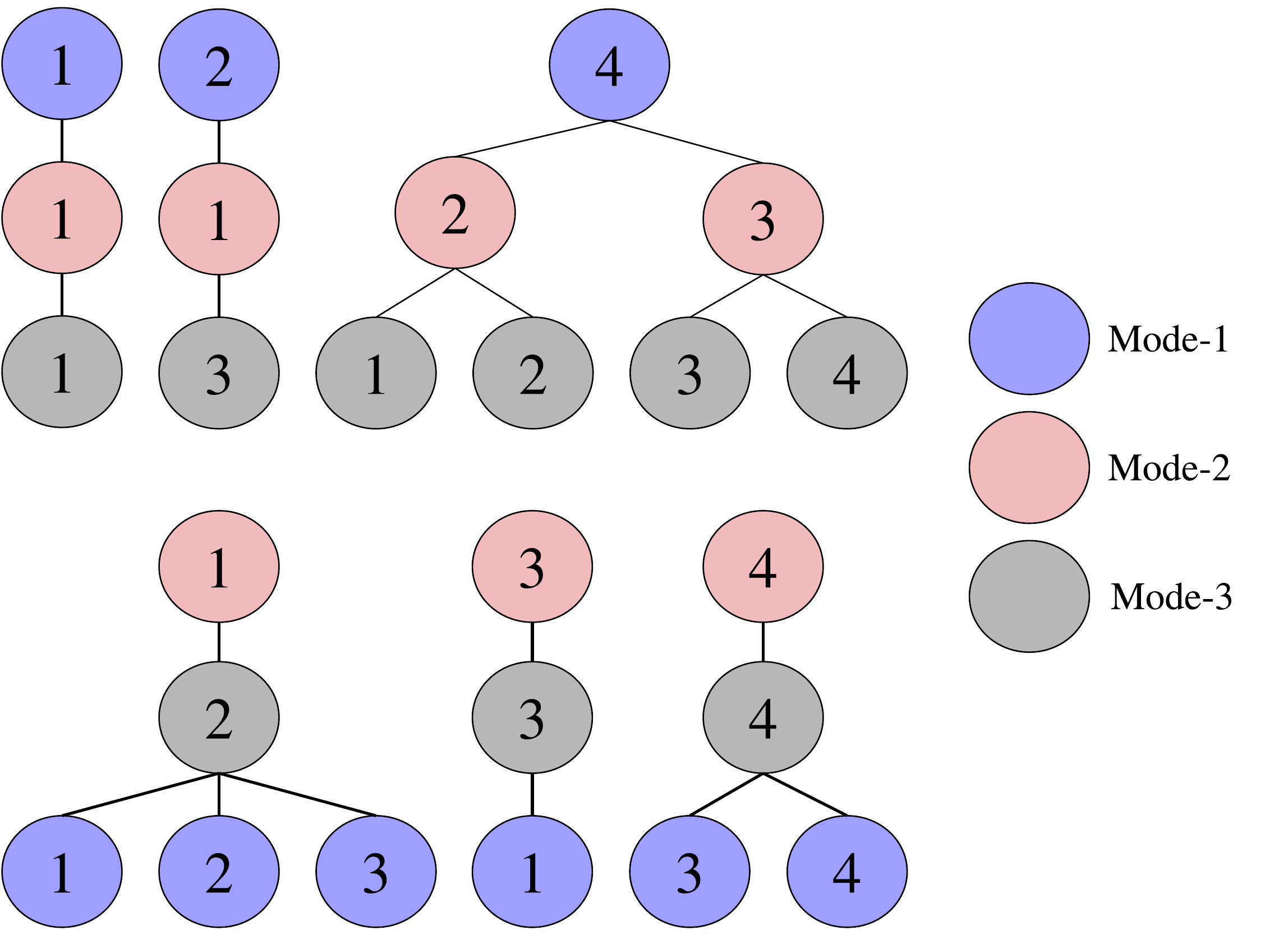}
    \vspace*{-5pt}
    \caption{MM-CSF format for the sparse tensor from Figure~\ref{fig:coo}.
    Unlike CSF, where every sub-tree has the same mode orientation, MM-CSF identifies fibers with the highest number of non-zero elements, and then constructs sub-trees 
    with different mode orientations.}
    \label{fig:mm-csf-fig}
    \vspace*{-8pt}
\end{figure}

Balanced CSF (B-CSF)~\cite{nisa2019load} creates sub-trees that are more balanced, but still requires $N$ copies of the tensor.
The state-of-the-art MM-CSF~\cite{nisa2019efficient} improves upon B-CSF by using a single copy of the tensor.
This is achieved by analyzing fiber density (i.e., number of non-zero elements in a fiber) and creating sub-trees with fibers that are as dense as possible.
However, the root of these sub-trees can come from any mode; hence, different traversal methods and parallel algorithms are required to perform MTTKRP, based on the mode that is being calculated, leading to drastic performance variations across different modes as shown in Figure~\ref{fig:motivation}.
Furthermore, the complexity of the tree-based structure requires different implementation for each tensor order (i.e., number of modes) and it also restricts MM-CSF to tensors that can fit in the limited GPU memory.
As a result, the current MM-CSF implementation only supports $3$- and $4$-dimensional tensors and cannot handle OOM tensors.
Figure~\ref{fig:mm-csf-fig} shows an example of the MM-CSF format for the tensor from Figure~\ref{fig:coo}.

\section{The BLCO Format}
\label{sec:blco}
To address the limitations of prior GPU-based formats, we propose the Blocked Linearized CoOrdinate (BLCO) format, a new sparse tensor representation devised for massively parallel architectures. BLCO linearizes and aggregates non-zero elements into coarse-grained blocks that meet the resource constraints of target GPUs to 
\begin{enumerate*}[label=(\roman*)]
    \item minimize data movement,
    \item accelerate indexing,
    \item enable a unified tensor representation and algorithmic implementation, and
    \item support out-of-memory tensor computation
\end{enumerate*}.
Figure~\ref{fig:blco-format} depicts an example of the BLCO format for the sparse tensor in Figure~\ref{fig:coo}. The generation of BLCO tensors consists of two stages: tensor linearization (Section~\ref{sec:blco-lin}) and adaptive blocking (Section~\ref{sec:blco-blocking}). 

\subsection{Tensor Linearization}\label{sec:blco-lin}
The BLCO format leverages index linearization~\cite{harrison2018high, ahmed2021} to map multi-dimensional space onto one-dimensional space, such that a point in N-D space, represented by $N$ coordinates, can be mapped to a point on an encoding line, represented by a \emph{single} index. The length of the encoding line determines the number of bits required to represent linearized indices. 
During computation (e.g., MTTKRP), the linear indices are \emph{de-linearized} to recover the original coordinates. Thus, fast de-linearization is important for high-performance execution of tensor algorithms using linearized formats.

\begin{figure}[tpb]
     \centering
     \begin{subfigure}[b]{0.2\textwidth}
         \centering
         \begin{tabular}{|rl|r|}
            \hline
             \multicolumn{2}{|c|}{$l$}  & \multicolumn{1}{|c|}{v} \\
             \hline
             0 & \small$(\textcolor{gray}{0}\textcolor{red}{0}\textcolor{blue}{0}\textcolor{gray}{0}\textcolor{red}{0}\textcolor{blue}{0})_{2}$ & 1.0\\
             \hline
             4 & \small$(\textcolor{gray}{0}\textcolor{red}{0}\textcolor{blue}{0}\textcolor{gray}{1}\textcolor{red}{0}\textcolor{blue}{0})_{2}$ & 2.0\\
             \hline             
             5 & \small$(\textcolor{gray}{0}\textcolor{red}{0}\textcolor{blue}{0}\textcolor{gray}{1}\textcolor{red}{0}\textcolor{blue}{1})_{2}$& 4.0\\
             \hline
             10 & \small$(\textcolor{gray}{0}\textcolor{red}{0}\textcolor{blue}{1}\textcolor{gray}{0}\textcolor{red}{1}\textcolor{blue}{0})_{2}$& 8.0\\
             \hline             
             12 & \small$(\textcolor{gray}{0}\textcolor{red}{0}\textcolor{blue}{1}\textcolor{gray}{1}\textcolor{red}{0}\textcolor{blue}{0})_{2}$& 6.0\\
             \hline
             15 & \small$(\textcolor{gray}{0}\textcolor{red}{0}\textcolor{blue}{1}\textcolor{gray}{1}\textcolor{red}{1}\textcolor{blue}{1})_{2}$& 9.0\\
             \hline             
             33 & \small$(\textcolor{gray}{1}\textcolor{red}{0}\textcolor{blue}{0}\textcolor{gray}{0}\textcolor{red}{0}\textcolor{blue}{1})_{2}$& 5.0\\
             \hline
             48 & \small$(\textcolor{gray}{1}\textcolor{red}{1}\textcolor{blue}{0}\textcolor{gray}{0}\textcolor{red}{0}\textcolor{blue}{0})_{2}$& 3.0\\
             \hline             
             57 & \small$(\textcolor{gray}{1}\textcolor{red}{1}\textcolor{blue}{1}\textcolor{gray}{0}\textcolor{red}{0}\textcolor{blue}{1})_{2}$& 10.0\\
             \hline             
             61 & \small$(\textcolor{gray}{1}\textcolor{red}{1}\textcolor{blue}{1}\textcolor{gray}{1}\textcolor{red}{0}\textcolor{blue}{1})_{2}$& 11.0\\
             \hline
             62 & \small$(\textcolor{gray}{1}\textcolor{red}{1}\textcolor{blue}{1}\textcolor{gray}{1}\textcolor{red}{1}\textcolor{blue}{0})_{2}$& 7.0\\             
             \hline
             63 & \small$(\textcolor{gray}{1}\textcolor{red}{1}\textcolor{blue}{1}\textcolor{gray}{1}\textcolor{red}{1}\textcolor{blue}{1})_{2}$& 12.0\\             
             \hline
         \end{tabular}
         \caption{Initial linearization.}
         \label{fig:lin}
     \end{subfigure}
     \hfill
     \begin{subfigure}[b]{0.25\textwidth}
         \centering
         \begin{tabular}{r|rl|r|}
            \cline{2-4}
             $b$ & \multicolumn{2}{c|}{$l$} & \multicolumn{1}{c|}{v} \\
             \hline
             \multirow{6}{*}{0} & 0 & \small$(\textcolor{gray}{0}\textcolor{red}{0}\textcolor{red}{0}\textcolor{blue}{0}\textcolor{blue}{0})_{2}$ & 1.0\\
             \cline{2-4}
             & 16 & \small$(\textcolor{gray}{1}\textcolor{red}{0}\textcolor{red}{0}\textcolor{blue}{0}\textcolor{blue}{0})_{2}$& 2.0\\
             \cline{2-4}             
             & 17 & \small$(\textcolor{gray}{1}\textcolor{red}{0}\textcolor{red}{0}\textcolor{blue}{0}\textcolor{blue}{1})_{2}$& 4.0\\
             \cline{2-4}
             & 6 & \small$(\textcolor{gray}{0}\textcolor{red}{0}\textcolor{red}{1}\textcolor{blue}{1}\textcolor{blue}{0})_{2}$& 8.0\\
            \cline{2-4}
             & 18 & \small$(\textcolor{gray}{1}\textcolor{red}{0}\textcolor{red}{0}\textcolor{blue}{1}\textcolor{blue}{0})_{2}$& 6.0\\
             \cline{2-4}
             & 23 & \small$(\textcolor{gray}{1}\textcolor{red}{0}\textcolor{red}{1}\textcolor{blue}{1}\textcolor{blue}{1})_{2}$& 9.0\\
             \hline             
             \multirow{6}{*}{1} & 1 &  \small$(\textcolor{gray}{0}\textcolor{red}{0}\textcolor{red}{0}\textcolor{blue}{0}\textcolor{blue}{1})_{2}$& 5.0\\
             \cline{2-4}             
             & 8 & \small$(\textcolor{gray}{0}\textcolor{red}{1}\textcolor{red}{0}\textcolor{blue}{0}\textcolor{blue}{0})_{2}$& 3.0\\
             \cline{2-4}  
             & 11 & \small$(\textcolor{gray}{0}\textcolor{red}{1}\textcolor{red}{0}\textcolor{blue}{1}\textcolor{blue}{1})_{2}$& 10.0\\
             \cline{2-4}             
             & 27 & \small$(\textcolor{gray}{1}\textcolor{red}{1}\textcolor{red}{0}\textcolor{blue}{1}\textcolor{blue}{1})_{2}$& 11.0\\
             \cline{2-4}
             & 30 & \small$(\textcolor{gray}{1}\textcolor{red}{1}\textcolor{red}{1}\textcolor{blue}{1}\textcolor{blue}{0})_{2}$& 7.0\\
             \cline{2-4}             
             & 31 & \small$(\textcolor{gray}{1}\textcolor{red}{1}\textcolor{red}{1}\textcolor{blue}{1}\textcolor{blue}{1})_{2}$& 12.0\\
             \hline
         \end{tabular}
         \caption{BLCO tensor.}
         \label{fig:blco}
     \end{subfigure}
        \vspace*{-10pt}
        \caption{
        BLCO format for the sparse tensor in Figure~\ref{fig:coo}, where the bits of linearized indices are color-coded to indicate different modes. First, BLCO linearizes the tensor (\ref{fig:lin}) based on a recursive partitioning of the multi-dimensional space~\cite{ahmed2021}. Next, it aggregates non-zero elements into blocks (\ref{fig:blco}) according to accelerators' resource constraints, namely, on-device memory, length of encoding line, and support for bit manipulation.
        For simplicity, we assume that the maximum length of the encoding line is $32$ ($2^{5}$) and each block has no more than $6$ non-zero elements. Note that BLCO re-encodes the linearized index ($l$) to allow the use of bitwise mask/shift (which are efficiently supported on GPUs) for de-linearization instead of bit-level gather/scatter. 
        }
        \label{fig:blco-format}
           \vspace*{-8pt}
\end{figure}

To efficiently encode multi-dimensional spaces with irregular shapes (which typically arise in higher-order sparse data) in a mode-agnostic way, prior state-of-the-art linear format (ALTO~\cite{ahmed2021}) recursively partitions the multi-dimensional space and ``traverses'' this space linearly using a compact space-filling curve. When the multi-dimensional space is regular, i.e., all modes have the same length, the resulting space-filling curve is similar to Morton-Z ordering~\cite{morton1966computer}. This linearization outperforms state-of-the-art tree-, list-, and block-based formats~\cite{ahmed2021} on CPU-based platforms, and therefore, we adopt its ordering for BLCO. 

Such a \emph{mode-agnostic} linearized encoding results in an \emph{interleaving} of bits from different mode indices, as shown in Figure~\ref{fig:lin}.
To quickly process indices encoded in this manner, efficient support for bit-level \emph{scatter} and \emph{gather} operations are needed.
However, GPUs lack native support for advanced bit manipulation, and emulating these instructions can be prohibitive,\footnote{For a third-order tensor, we estimate that a na\"ive emulation would require $276$ bitwise operations to de-linearize each non-zero element.} leading to inefficient tensor processing operations. 

To address this issue, the BLCO format arranges the non-zero elements using ALTO ordering, but \emph{re-encodes} the linearized indices to allow the use of bitwise \emph{shift} and \emph{mask} (AND) instructions, which are natively supported on accelerators, to de-linearize the indices. As illustrated in Figure~\ref{fig:blco}, BLCO achieves this goal by rearranging the encoding bits of linearized indices ($l$) into contiguous mode sets that can be quickly extracted on GPUs during tensor computations. This re-encoding resembles the index concatenation of the LCO \cite{harrison2018high} format, but with bitwise shift and mask used for de-linearization instead of arithmetic division and modulo. 

\subsection{Adaptive Blocking}\label{sec:blco-blocking}
To map a tensor \TENSOR{X} $\in$ \REALX{I}{N} into a linearized form, an encoding line of length $I_{1}\times \cdots \times I_{N}$ is required.
As the number of modes and their lengths increase, the encoding line and the range of linear index values can significantly expand.
Hence, large-scale tensors may require more than $64$ bits to encode a linear index.
Since GPUs do not provide native support for large integer (more than $64$ bits) operations, a custom implementation\footnote{ https://github.com/curtisseizert/CUDA-uint128} of large integer arithmetic is needed, which are typically not as efficient as native instructions.
In addition, many tensors can require a slightly higher bit resolution (only a \emph{few} additional bits) than $64$ bits, so using large integers for linearized indices may lead to wasted memory.

Most importantly, the state-of-the-art libraries for sparse MTTKRP require the \emph{entire} tensor to be in the \emph{limited} GPU memory, due to their compressed and/or mode-specific tensor formats. In particular, it is challenging to stream the data in small \emph{chunks} that can be partially processed using tree- (e.g., B-CSF~\cite{nisa2019load} and MM-CSF~\cite{nisa2019efficient}) and block-based (e.g., HiCOO~\cite{li2018hicoo}) formats, where the granularity of tensor chunks (e.g., sub-trees and HiCOO blocks) are difficult to control.
List-based formats (e.g., GenTen~\cite{phipps2019software} and F-COO~\cite{liu2017unified}) are more amenable to streaming, but will require additional information for synchronization across these data chunks.

To address these issues, we propose \emph{adaptive blocking} to aggregate non-zero elements into coarse-grained blocks by partitioning the tensor into smaller sub-tensors.
Blocking the tensor to exploit density structures and to compress the data has been previously explored by the HiCOO~\cite{li2018hicoo} format. 
However, such a compression comes at the cost of severe workload imbalance across blocks, due to the irregular spatial distributions of sparse data~\cite{ahmed2021, li2019efficient}, and it requires expensive tuning to find the best block size. 
In contrast, the proposed blocking technique aims to meet the requirements of memory-constrained GPUs, while at the same time generate the largest possible blocks that can efficiently utilize these throughput-oriented accelerators with massive parallelism.

As such, our BLCO format first uses the \emph{uppermost bits} from every mode of linearized indices that exceeds the target integer size to form the initial blocks, which allows the sub-spaces spanned by these blocks to naturally adapt to the underlying tensor space. 
For example, if a tensor requires $72$ bits for linearized indices and the size of target integers is $64$ bits, a total of $8$ ($72 - 64$) 
uppermost bits across all the modes 
are stripped from the linear indices and used as a \emph{key} to group the non-zero elements into blocks; then, these bits are stored as metadata along with each block ($b$), as depicted in Figure~\ref{fig:blco}.
This strategy does \emph{not} require expensive tuning to create the blocks and, compared to using longer encoding lines, it reduces the memory footprint and leverages more efficient native integer ($64$-bit) instructions. Next, BLCO further splits the initial blocks, if needed, based on the available on-device memory to ensure that each block has no more than the maximum number of non-zero elements that can fit in the target device.

While the resulting BLCO format may still suffer from variance in the number of non-zero elements across blocks, it exposes the fine-grained parallelism within a block, due to its linearized list-based form. 
Specifically, it allows workload to be partitioned at the granularity of a non-zero element rather than a compressed block. 
Hence, a GPU hardware scheduler can automatically hide the execution and memory-access latency as well as balance the workload across threads, as long as  there are enough non-zero elements to be processed in the GPU memory. 
In addition, our massively parallel MTTKRP algorithm (Section~\ref{sec:hreduction}), with opportunistic conflict resolution, allows BLCO blocks to be processed independently. Therefore, once the GPU has processed a block, it will fetch the next available block to automatically occupy any available GPU resources. These properties enable our BLCO format and MTTKRP algorithm to seamlessly handle OOM tensors, in contrast to prior work.

Our implementation uses device queues (SYCL queues or CUDA streams) to launch BLCO blocks, allowing the computation of active blocks to be done asynchronously with the transmission of pending blocks. 
Each device queue has reserved memory, corresponding to the number of non-zero elements that can be processed at one time, which is reused across BLCO blocks assigned to the same queue.
In our experiments, we use up to $8$ device queues and set the maximum number of non-zero elements per block to $2^{27}$ to fill the GPU; these parameters allow for enough concurrency and further tuning them has negligible performance impact.

Hypersparse tensors may generate several BLCO blocks that can fit in the same memory reservation of a single device queue. 
Launching these blocks across multiple queues can potentially incur kernel launch overhead on some GPU architectures. To address this, we batch all BLCO blocks that can be processed by one device queue into a GPU kernel and explicitly store block mappings and element offsets at the work-group (thread block) boundaries. This additional batching information is calculated during format construction and thus creates minimal overhead during tensor computations.

\section{BLCO-based Sparse MTTKRP}
\label{sec:hreduction}
On massively parallel systems such as GPUs, atomic updates to global memory can be expensive because of the sheer number of concurrently executing threads that could conflict and the long data access latency.
Therefore, prior studies focused on reducing atomic operations by storing the data in a mode-specific form~\cite{nisa2019efficient,nisa2019load, Liu2017, Dun2021} and/or keeping mode-specific mapping/scheduling information~\cite{phipps2019software, Liu2017, Dun2021}.
However, these strategies lead to drastic performance variation across modes, substantial memory overhead for extra tensor copies, and/or complex algorithms and implementations (c.f. Section~\ref{sec:challenges}).

\begin{figure}[tb]
    \centering
    \includegraphics[width=1.0\linewidth]{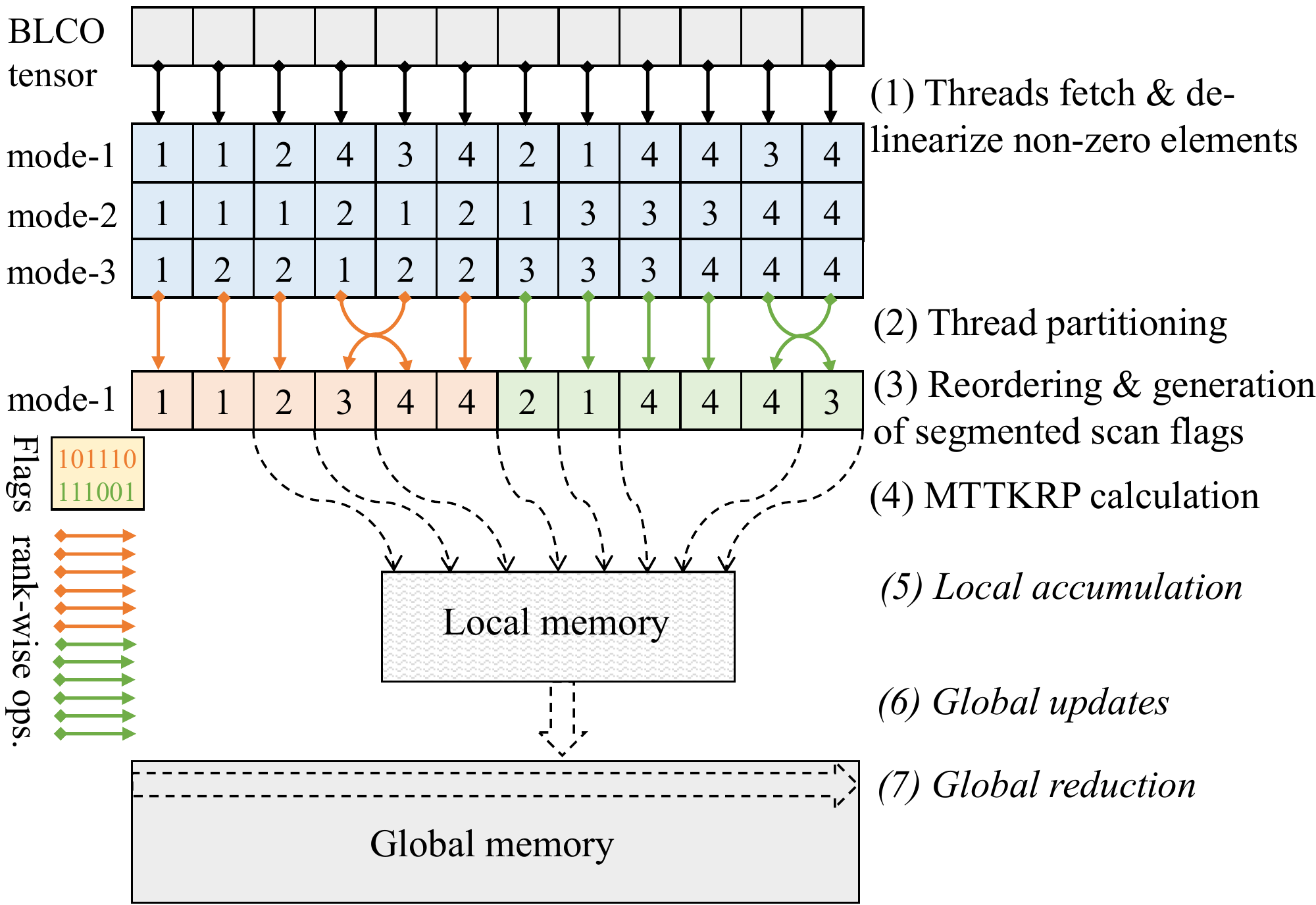}
    \vspace*{-20pt}
    \caption{
    Steps (1) -- (7) illustrates mode-$1$ MTTKRP with hierarchical conflict resolution for the BLCO tensor from Figure~\ref{fig:blco}.
    For register-based conflict resolution, step (6) is the last one, where the results are written directly, bypassing the local accumulation in step (5), to the factor matrix in global memory using atomic updates, as opposed to writing to temporary factor matrix copies.
    In all conflict resolution mechanisms, steps (1) -- (4) are common and writing the updates happens at segment boundaries.}
    \vspace*{-10pt}
    \label{fig:reduce}
\end{figure}

Here, we describe a novel massively parallel MTTKRP algorithm that eliminates control-flow and memory-access irregularities while resolving update conflicts (RAW hazards) using an \emph{opportunistic on-the-fly}
update mechanism, which reduces atomic operations without requiring extra tensor copies or mode-specific information.

\subsection{Hierarchical Conflict Resolution}\label{blco-h}
Figure~\ref{fig:reduce} illustrates our massively parallel algorithm for mode-$1$ MTTKRP kernel on the third-order sparse tensor from Figure~\ref{fig:coo}.
Note that the non-zero elements are linearized and grouped according to the BLCO format (Figure~\ref{fig:blco}).
For simplicity, we assume a work-group (thread-block) size of $12$ and tile size of $6$ work-items (threads) in the figure.

To eliminate control-flow and memory-access irregularities, the proposed algorithm has two phases: processing and computing.
In the processing phase, steps (1) -- (3),
each thread is assigned to a non-zero element
and threads collaborate to perform \emph{on-the-fly} de-linearization and reordering of non-zero elements as well as generation of segmented scan flags. 
In the computing phase, steps (4) -- (7), threads are reassigned to perform the rank-wise MTTKRP computations and merge conflicting updates at the register, local memory, and global memory levels. 
In each phase, the threads and their global memory accesses are coalesced.    
\subsubsection{Processing Phase}
In step (1), a group of threads (work-group or thread-block) is assigned to a certain number of non-zero elements and they first collaborate so that each thread loads a linearized non-zero element (i.e., a linear index and its corresponding value) from the global memory in a coalesced manner.
Each thread then proceeds to de-linearize the linear index and recover the multi-dimensional coordinates, where each coordinate can be calculated independently to expose more instruction-level parallelism.
In step (2), the threads are partitioned into multiple tiles, where a tile size is no more than the sub-group (warp) size. Next, in step (3), the threads within a tile collaborate to reorder the non-zero elements into groups (segments) according to their target mode indices (mode-$1$ in the example) and then generate segmented scan flags, where $1$ indicates the start of a new segment.
To minimize thread divergence and data access latency, we implement the reordering of non-zero elements via parallel histogram and prefix sum, using sub-group (warp-level) data exchange primitives (e.g., \emph{shuffle} operations), and broadcast/store the segmented scan flags across threads in a low-latency register. After each thread completes the on-the-fly processing phase, the non-zero elements are stored, according to their new order, in local memory for later access.

\subsubsection{Computing Phase}
Once the data has been processed, we reassign the threads to calculate the rank-wise MTTKRP computations in step (4), where each thread is responsible for one or more elements along the decomposition rank (i.e., threads process the same non-zero element and perform rank-wise operations in parallel).
The algorithm iterates over the non-zero elements using the segmented scan flags, so that each thread accumulates its partial results to a register as long as the target mode index remains the same.
At the end of each segment, i.e., when the index changes, each thread writes the accumulated result to the stash (a software-controlled cache) in local memory (shown as ``step (5)'' in the figure).
In step (6), when all the non-zero elements have been processed, the results are copied from the stash (local memory) to one of the multiple copies of the factor matrix in global memory.
By using multiple copies of the factor matrix, we can minimize the probability of conflict when multiple thread blocks are copying their results back to global memory at the same time.
Finally, in step (7), the multiple factor matrix copies are merged in global memory to produce the final result.

\subsection{Register-based Conflict Resolution}
Our register-based conflict resolution is a subset of the hierarchical algorithm, discussed above.
It bypasses the local memory entirely and writes the accumulated result in registers to global memory, without the need for a final global reduction.
Specifically, after each thread calculates and accumulates the result in its register and reaches a segment boundary, it bypasses step (5) and completes execution at step (6) after writing  
the updates to the final factor matrix, as opposed to one of its copies, using atomic operations.

\subsection{Adaptation Heuristic}\label{blco-adaptive}
The proposed synchronization mechanisms use different number of atomic operations to perform MTTKRP.
Register-based conflict resolution requires more atomic operations, as updates at each segment boundary need atomic add operations to the factor matrix in global memory.
Hierarchical conflict resolution uses fewer atomic operations, as updates at each segment boundary are copied first to the local-memory stash, and then atomic operations are used only at the end of the work-group (thread-block) execution to write the accumulated result in the stash to the factor matrix in global memory.
Furthermore, multiple copies of the factor matrix can be used to reduce the probability of an update conflict, at the cost of a final global reduction.

We propose a simple heuristic for selecting the best conflict resolution mechanism based on the characteristics of target modes and GPU devices. In modern GPUs~\cite{blythe2020xe, nvidia2021}, the execution units (EUs) are aggregated into subslices or streaming multi-processors (SMs). Multiple subslices are grouped into a GPU slice or a graphics processing cluster (GPC).
Our heuristic selects the hierarchical conflict resolution, when the target mode length is less than the number of subslices (SMs). 
At this mode length, the contention from atomic operations to global memory is severe, and using a local-memory stash and factor matrix copies (one copy for each slice or GPC) alleviates this contention.  
For all other cases, we use the register-based conflict resolution.

\section{Experiments}
\label{sec:exp}
We evaluate the proposed sparse MTTKRP and BLCO format\footnote{Available at https://github.com/jeewhanchoi/blocked-linearized-coordinate} using a representative set of in-memory and out-of-memory tensors with different characteristics. We compare our performance to the state-of-the-art sparse tensor frameworks for massively parallel GPUs, namely, 
MM-CSF~\cite{nisa2019efficient}, GenTen~\cite{phipps2019software}, and F-COO~\cite{Liu2017}. While the other frameworks only support tensors that can fit in the device memory, BLCO can process both in- and out-of-memory tensors and delivers superior performance across all in-memory tensors.

\begin{table}[t]
\small 
    \centering
	\caption{Hardware and software setup.}
	\vspace*{-10pt}
	\begin{threeparttable}
	\begin{tabular}{|l|l|l|l|}
		\hline
		& \multicolumn{1}{c|}{CPU} & \multicolumn{2}{c|}{GPU} \\
		\hline
        Model & AMD  & NVIDIA  & NVIDIA \\
        & EPYC 7662 & A100 & V100\\
        \hline
        $\mu$-arch & Zen 2 & Ampere & Volta \\
        \hline
		Frequency & 3.3 GHz & 1.41 GHz & 1.38 GHz \\
		\hline
        Cores & 64 ($\times$2 sockets) & 108 (SM\tnote{1}\ \ ) & 80 (SM) \\
        & & 6912 (CC\tnote{2}\ \ ) & 5120 (CC) \\
        \hline
        Caches & 2MB L1,  & 15MB L1D, & 10MB L1D, \\
        & 32MB L2, & 40MB L2 & 6MB L2\\
        & 256MB L3 & & \\
        \hline
		DRAM & 256 GB & 40 GB & 32 GB \\
		(Bandwidth) & & (1555 GB/s) & (900 GB/s)\\
		\hline
        Interconnect & PCI-e Gen 4 & NVLink 3.0 & NVLink 2.0\\
		\hline
        OS/Driver & RHEL & Driver  & Driver \\
        (version) & (8.3) & (470.42.01) & (440.33.01) \\
        \hline
		Compiler & gcc 9.3.0 & nvcc 11.4 & nvcc 11.0 \\
		\hline
	\end{tabular}
	\begin{tablenotes}\footnotesize
    \item[1] Streaming Multiprocessor\qquad $^{2}$ CUDA Cores
    \end{tablenotes}
	\end{threeparttable}
	\vspace*{-5pt}
	\label{tab:hardware}
\end{table}

\subsection{Evaluation Setup}
\subsubsection{Test Platform}
We conduct the experiments on the latest Intel discrete GPU with a single-tile (denoted \emph{Intel Device1}) and two NVIDIA GPUs from the most recent micro-architecture generations: A100 (\emph{Ampere}) and V100 (\emph{Volta}).  
For the format generation, we use a dual-socket AMD Epyc 7662 CPU system and employ $128$ threads. 
Our prototype is implemented in Data Parallel C++ (DPC++)~\cite{reinders2021data} as well as CUDA~\cite{kirk2006nvidia}. Since the current sparse tensor frameworks lack portability across GPU architectures, we ported the state-of-the-art MM-CSF to DPC++ using the Intel DPC++ Compatibility Tool~\cite{dpct}.
Due to confidentiality requirements, Table~\ref{tab:hardware} summarizes the publicly available specifications of the target hardware and software environment.

\begin{table}[tb]
\small  
\centering
\caption{The sparse tensor data sets used for evaluation,
ordered by the number of non-zero elements.}
\vspace*{-10pt}
\label{tab:problems}
\begin{tabular}{|p{1.2cm}|p{3.25cm}|p{0.85cm}|p{1.18cm}|p{1.2cm}|}
\hline
Tensor & Dimensions & NNZs & Density \\
\hline
    NIPS & $ 2.5K\times 2.9K\times 14K\times17$ & $3.1M$ & $1.8\times10^{-06}$\\\hline 
    Uber & $ 183\times24\times 1.1K\times1.7K$&  $3.3M$ & $3.8\times10^{-04}$\\\hline 
    Chicago & $ 6.2K\times24\times77\times32$& $5.3M$& $1.5\times10^{-02}$\\\hline 
    Vast-2015 & $165.4K\times11.4K\times2$& $ 26M$& $7.8\times10^{-07}$\\\hline 
    DARPA & $22.5K\times22.5K\times23.8M$& $28.4M$& $2.4\times10^{-09}$\\\hline 
    Enron & $ 6K\times5.7K\times244.3K\times1.2K$ & $54.2M$& $5.5\times10^{-09}$\\\hline 
    NELL-2 & $12.1K\times9.2K\times28.8K$& $76.9M$& $2.4\times10^{-05}$\\\hline 
    FB-M & $23.3M\times23.3M\times166$ & $99.6M$& $1.1\times10^{-09}$\\\hline 
    Flickr & $319.7K\times28.2M\times1.6M\times731$& $112.9M$& $1.1\times10^{-14}$\\\hline 
    Delicious & $532.9K\times17.3M\times2.5M\times1.4K$& $140.1M$& $4.3\times10^{-15}$\\\hline 
    NELL-1 & $2.9M\times2.1M\times25.5M$& $143.6M$& $9.1\times10^{-13}$\\\hline 
    Amazon & $4.8M\times1.8M\times1.8M$& $1.7B$ & $1.1\times10^{-10}$\\\hline 
    Patents & $46\times239.2K\times239.2K$& $3.6B$ & $1.4\times10^{-03}$\\\hline 
    Reddit & $8.2M\times177K\times8.1M$& $4.7B$& $4.0\times10^{-10}$\\\hline 
\end{tabular}
\label{tab:tensor-data}
\vspace*{-5pt}
\end{table}

\begin{figure*}[tpb]
    \centering
    \includegraphics[width=1.0\linewidth]{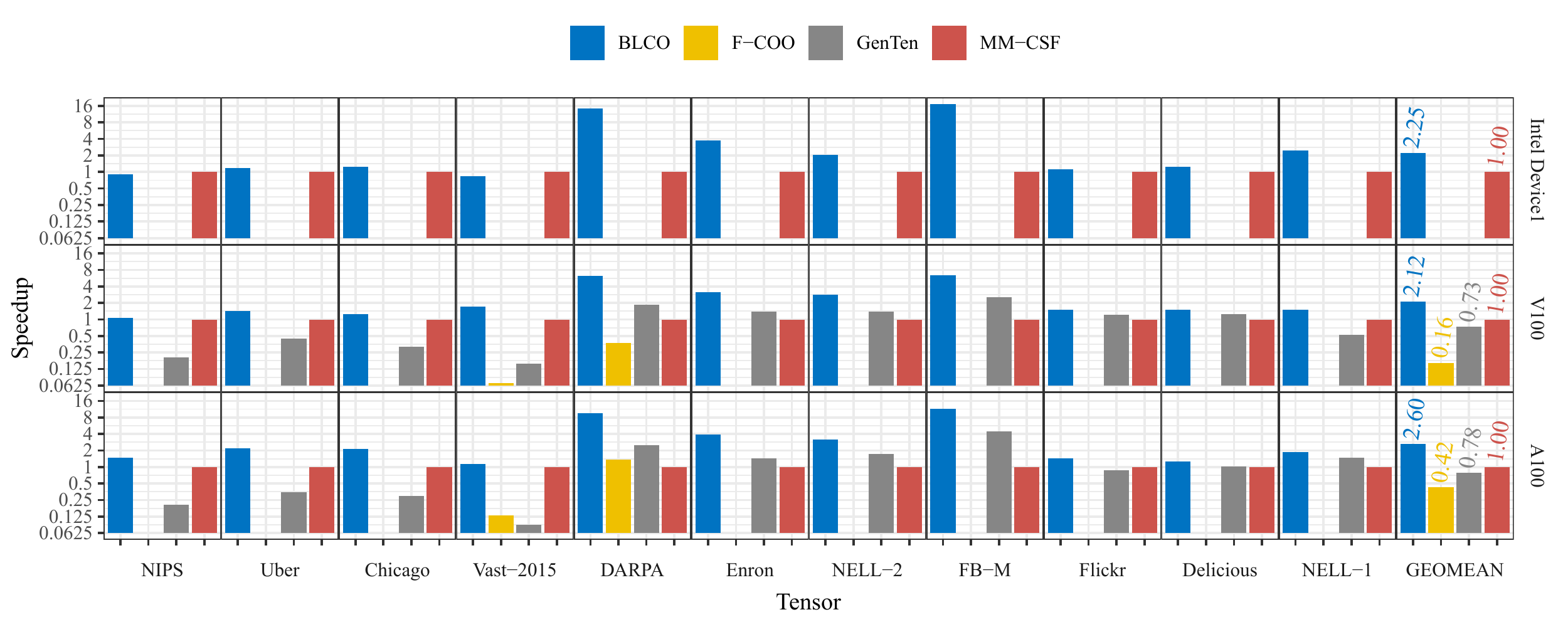}
	\vspace*{-24pt}
	\caption{Comparison of BLCO-based MTTKRP against popular GPU frameworks on the data sets that fit in GPU memory.
	Each bar represents the speedup obtained 
	against MM-CSF for computing MTTKRP on all modes.
	The right-most group of bars show the geometric mean speedup.
	}
   \vspace*{-2pt}
\label{fig:framework-comparison}
\end{figure*}

\subsubsection{Data Sets}
We consider $14$ \emph{real-world} tensor 
data sets 
from the FROSTT~\cite{frostt} and HaTen2~\cite{haten2_ICDE2015} open-source repositories that cover a wide range of tensor properties and sparsity structures---number of modes, mode lengths, number of non-zero elements, and density.
Table~\ref{tab:tensor-data} lists the 
sparse tensor data sets used for evaluation, ordered by increasing number of non-zero elements (NNZs). The large-scale tensors with billions of non-zero elements (namely, Amazon, Patents, and Reddit) are considered out-of-memory as they fail to execute on current tensor decomposition frameworks, which need to keep the entire tensor and factor matrices in device memory, producing memory allocation errors on target GPUs.

\subsubsection{Configurations}
The experiments use a decomposition rank of 32 for MTTKRP, as per prior work~\cite{nisa2019efficient}. Using double-precision values and 64-bit integers, we report the performance as an average over $25$ iterations. We tune each sparse tensor framework to the best of our abilities. For the state-of-the-art MM-CSF, we exhaustively tune the number of warps per fiber and the thread-block size and report the best execution time. 
While BLCO can benefit from tuning, we use our adaptation heuristic (Section~\ref{blco-adaptive}) and set the thread-coarsening factor (NNZs per thread) for the hierarchical conflict resolution mechanism (Section~\ref{blco-h}) to $4$ and $2$ on Intel and NVIDIA GPUs, respectively. 
\begin{figure*}[tpb]
    \centering
    \includegraphics[width=1.0\linewidth]{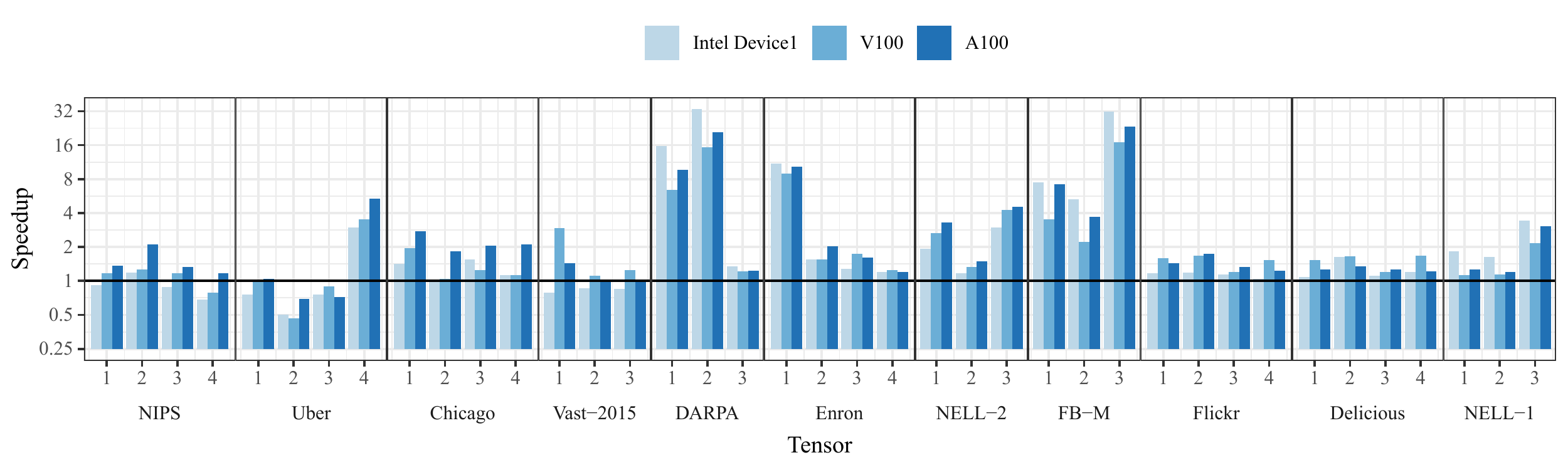}
	\vspace*{-24pt}
	\caption{The per-mode speedup of BLCO-based MTTKRP against MM-CSF across \emph{every} tensor mode for the data sets that fit in GPU memory.
	}
   \vspace*{-4pt}
\label{fig:speedup-mmcsf}
\end{figure*}

\subsection{Comparison Against TD Frameworks}
We first compare our BLCO-based MTTKRP against the other popular sparse tensor decomposition frameworks.
Figure \ref{fig:framework-comparison} shows the MTTKRP execution time for all modes, across the GPU frameworks, normalized by the execution time of the state-of-the-art MM-CSF.
The results demonstrate that BLCO consistently outperforms all other frameworks, achieving a geometric mean speedup between $2.12\times$ and $2.6\times$ over MM-CSF across the different GPU devices.
For the other CUDA frameworks, GenTen has comparable performance to MM-CSF, outperforming MM-CSF on six out of $11$ data sets, whereas F-COO has lower performance on average compared to MM-CSF, especially on the V100 GPU. 
The missing data points for F-COO is due to its limited support for higher-order tensors (i.e., 3-D only) and "segfault" errors.

Our performance analysis indicates that the performance of MM-CSF (as well as GenTen) can substantially decrease with higher synchronization cost, which leads to lower average performance than BLCO on the GPU devices with more expensive synchronization. 
Since MM-CSF reorders non-zero elements to increase tensor compression, its performance is sensitive to the number of non-zero elements per fiber. On large-scale data sets with low fiber density, such as DARPA, Enron, and FB-M, MM-CSF has lower compression, leading to significant performance degradation compared to BLCO.     

\subsection{Comparison Against State of the Art}
For a comprehensive evaluation against the state-of-the-art MM-CSF, Figure~\ref{fig:speedup-mmcsf} demonstrates the speedup achieved by our BLCO-based MTTKRP for every mode of the tensors that can fit in the device memory of target GPUs.
The results demonstrate that our BLCO format achieves better or comparable performance to MM-CSF for every mode (up to $33.35\times$ speedup) across all data sets, except Uber and NIPS.
These data sets are not only small, but they also have exceptionally short modes, allowing the data to fit in cache; thus, the higher compression achieved by MM-CSF, due to its mode-specific nature (c.f. Table~\ref{tab:mem}), translates to better performance. 
Yet, such a mode-specific compression leads to substantial performance variations across different modes, as shown in Figure~\ref{fig:motivation}, and as a result, BLCO still outperforms MM-CSF for all-mode MTTKRP (c.f. Figure~\ref{fig:framework-comparison}). 

\subsection{Memory Traffic Analysis}
Since sparse tensor decomposition is a memory-bound workload, its performance is largely limited by the data volume and the effective memory throughput. Hence, we provide detailed analysis of these memory metrics across both in- and out-of-memory tensors.

\subsubsection{In-Memory Tensors}
Table~\ref{tab:mem} details the memory metrics of BLCO-based MTTKRP compared to MM-CSF on the A100 GPU. 
The metrics are collected using the Nsight Compute profiler~\cite{nsight-compute}.
The memory analysis shows that MM-CSF achieves higher compression than BLCO thanks to its tree-like data structure, and the total volume of data fetched from memory (as shown in column ``Vol'') is lower in most cases.
However, due to the irregular memory access and expensive synchronization associated with traversing a tree-like tensor representation, MM-CSF under-utilizes the memory system compared to BLCO and has lower memory throughput (as shown in column ``TP'').
In addition, both the memory volume and throughput of MM-CSF vary significantly across modes because of its mode-specific traversal and processing of tensors, which leads to substantial performance variations (c.f. Figure~\ref{fig:motivation}).  
In contrast, while BLCO requires more data volume because of its mode-agnostic form, it achieves higher memory throughput by eliminating memory-access irregularities, exploiting data locality, and merging conflicting updates across threads in low-latency registers and memories. Thereby, BLCO fetches/writes more data from/to higher levels of the memory hierarchy in a coalesced way, leading to improved performance by up to an order of magnitude compared to MM-CSF.

\begin{table}[htb]
    \centering
    \caption{Comparison of the memory related metrics between BLCO and MM-CSF for MTTKRP on the A100 GPU.}
    \vspace*{-8pt}
    \resizebox{\columnwidth}{!}{
	\begin{threeparttable}

    \begin{tabular}{|c|c|l|l|l||c|c|l|l|l|}
    
        \hline
        \multirow{1}{*}{Data Set} & \multirow{1}{*}{Format} & \multirow{1}{*}{$n$} & Vol$^{1}$ & TP$^{2}$ & \multirow{1}{*}{Data Set} & \multirow{1}{*}{Format} & \multirow{1}{*}{$n$} & Vol$^{1}$ & TP$^{2}$\\
        \hline

        \multirow{8}{*}{Uber} & \multirow{4}{*}{BLCO} & 1 & 2.78 & 3.60 & \multirow{8}{*}{Enron} & \multirow{4}{*}{BLCO}
          & 1 & 44.82 & 4.11 \\
        & & 2 & 2.75 & 3.61 & & & 2 & 46.23 & 4.62\\
        & & 3 & 2.75 & 3.53 & & & 3 & 47.88 & 4.92\\
        & & 4 & 2.73 & 2.77 & & & 4 & 47.22 & 4.70\\
         \cline{2-5} \cline{7-10}
 
        & \multirow{4}{*}{MM-CSF} & 1 & 1.68 & 1.68 &  & \multirow{4}{*}{MM-CSF} & 1 & 41.39 & 0.31 \\
        & & 2 & 1.33 & 2.03 & & & 2 & 62.83 & 3.16\\
        & & 3 & 1.33 & 1.93 &  & & 3 & 37.15 & 2.29\\
        & & 4 & 2.12 & 0.32 & & & 4 & 37.05 & 3.01\\
         \hline
        
        \multirow{6}{*}{Vast-2015} & \multirow{3}{*}{BLCO} & 1 & 16.91 & 3.92 & \multirow{6}{*}{NELL-1} & \multirow{3}{*}{BLCO} & 1 & 107.5 & 2.44 \\
        & & 2 & 16.73 & 3.77 & & & 2 & 104.5 & 2.32\\
        & & 3 & 13.92 & 2.90 & & & 3 & 110.7 & 2.39\\
        \cline{2-5} \cline{7-10}
 
        & \multirow{3}{*}{MM-CSF} & 1 & 9.19 & 1.19 &  & \multirow{3}{*}{MM-CSF} & 1 & 123.1 & 2.21 \\
        & & 2 & 8.36 & 1.57 & & & 2 & 118.5 & 2.19 \\
        & & 3 & 8.36 & 1.45 & & & 3 & 122.1 & 0.86\\
        \hline
    \end{tabular}
    \begin{tablenotes}
    \item[1] Memory volume in GB, measured by \emph{l1tex\_\_t\_bytes.sum} in Nsight Compute \cite{nsight-compute}
    \item[2] Memory throughput in TB/s, calculated by (Vol / total execution time)
    \end{tablenotes}
	\end{threeparttable}
}
	\vspace*{-5pt}
    \label{tab:mem}
\end{table}

\subsubsection{Out-of-Memory Tensors}
\begin{figure}[tpb]
    \centering
    \includegraphics[width=1.0\linewidth]{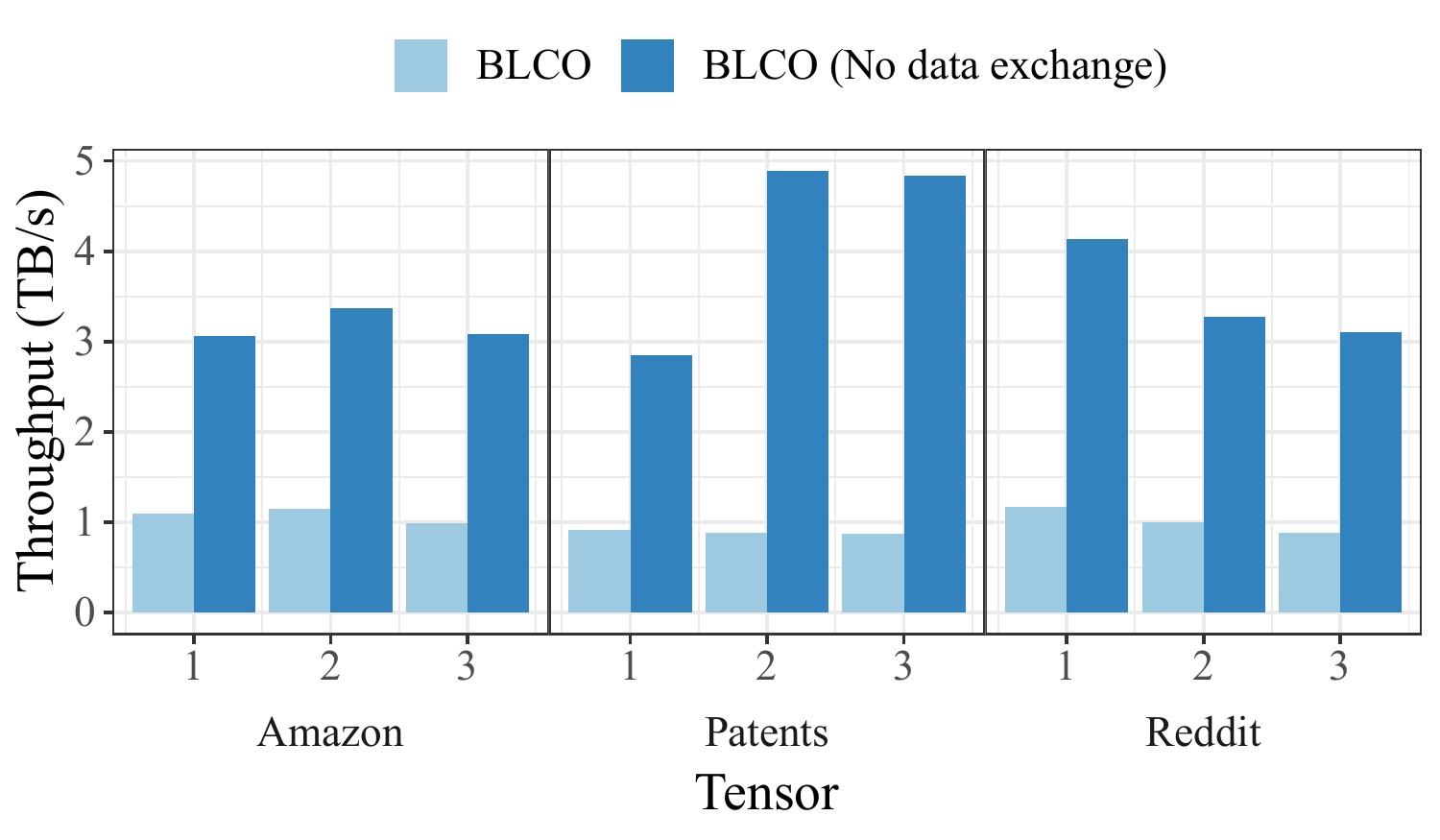}
	\vspace*{-24pt}
	\caption{The memory throughput of BLCO-based MTTKRP (with and without host-device data exchange) for out-of-memory tensors across every mode on the A100 GPU.}
   \vspace*{-6pt}
\label{fig:oom-tensor}
\end{figure}

Figure \ref{fig:oom-tensor} demonstrates the memory throughput of BLCO-based MTTKRP for out-of-memory tensors (Amazon, Patents, and Reddit) on the A100 GPU.
Since no other GPU framework supports these tensors, a direct comparison is not possible; 
instead, we report the overall throughput as well as the throughput without host-device data exchange (in-memory throughput) of our BLCO-based MTTKRP, as measured using the Nsight Systems profiler \cite{nsight-systems}.
The overall throughput is measured based on the total execution time (both MTTKRP computations and host-device data transfers), while the in-memory throughput only includes MTTKRP computations. 
Without the overhead of host-device communication resulting from the limited GPU memory, the in-memory throughput is on par with the performance observed for the  tensors that can fit in the GPU (c.f. Table \ref{tab:mem}).
However, the overall throughput is lower due to the limited bandwidth of the host-device interconnect compared to the device memory bandwidth.
While BLCO achieves perfect overlap between host-device transfers and MTTKRP computations, the communication overhead still dominates the execution time, leading to lower overall throughput ($57\%$--$75\%$ of the memory bandwidth).
This result demonstrates that our framework handles out-of-memory tensors as efficiently as in-memory tensors, but the overall performance for out-of-memory tensors is limited by the bandwidth of the host-device interconnect.
 
\subsection{Format Construction}
\begin{figure*}[htb]
    \centering
    \includegraphics[width=1.0\linewidth]{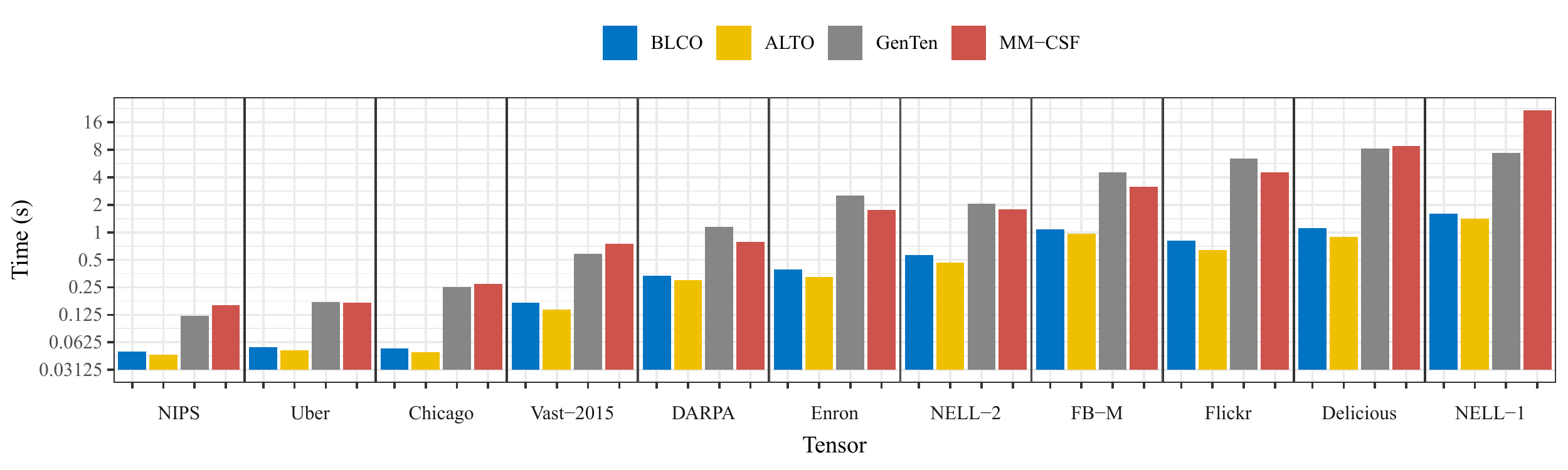}
	\vspace*{-24pt}
 	\caption{Comparison of the BLCO construction/generation cost against popular GPU formats as well as the CPU-based ALTO format on the data sets that fit in GPU memory.}
   \vspace*{-2pt}
\label{fig:preprocessing}
\end{figure*}

\begin{figure*}[htb]
    \centering
    \includegraphics[width=1.0\linewidth]{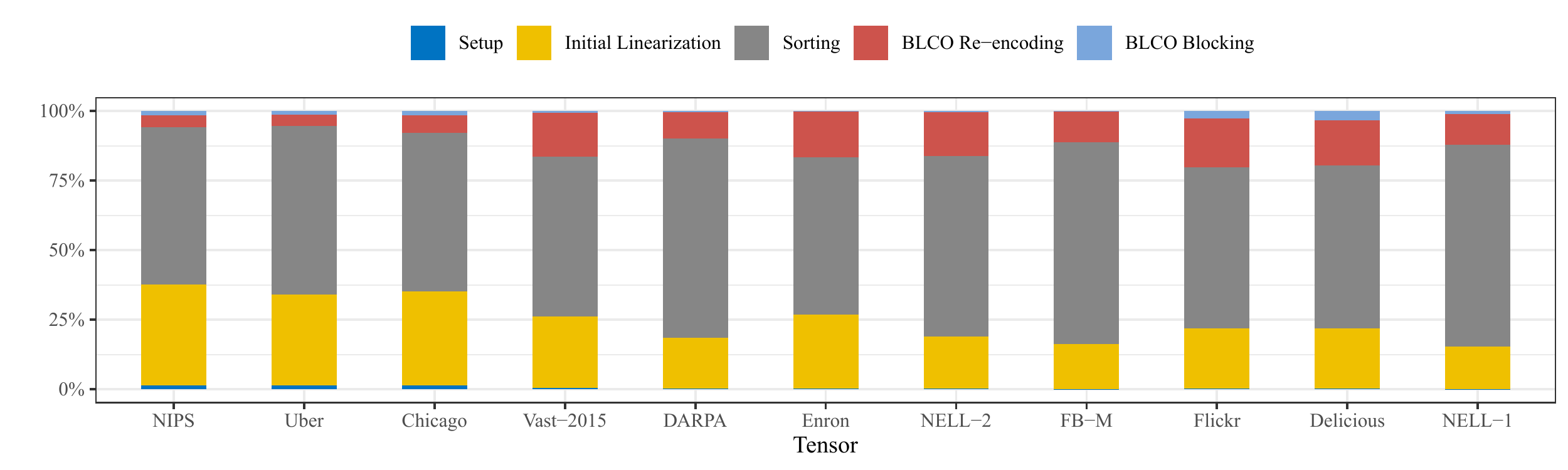}
	\vspace*{-24pt}
	\caption{Breakdown of the BLCO construction cost for the data sets that fit in GPU memory. Compared to ALTO, BLCO requires additional blocking and re-encoding to enable efficient execution on GPUs.}
\label{fig:breakdown}
\end{figure*}

Figure \ref{fig:preprocessing} shows the generation time of the GPU sparse formats, namely, BLCO, GenTen, and MM-CSF, as well as the CPU-based ALTO format.
Furthermore, Figure~\ref{fig:breakdown} details the run-time distribution across the different format generation stages of BLCO.
The tensor formats are generated from raw data, in the COO representation, on the host CPU listed in Table~\ref{tab:hardware}.
By adopting a mode-agnostic linearized form rather than a multi-dimensional representation, BLCO significantly decreases the format generation cost, which is typically dominated by sorting and clustering non-zero elements. 
Additionally, BLCO does not require extra mode-specific mapping or scheduling information for reducing synchronization.
As a result, BLCO is several times (up to $13.6\times$) cheaper to generate than the state-of-the-art MM-CSF format. On the A100 GPU, BLCO needs approximately $12$ full (all-mode) MTTKRP iterations on average to amortize its format generation time, while the other GPU formats require up to an order of magnitude more iterations to amortize their construction cost.  

Compared to ALTO, BLCO enables efficient execution of tensor operations on massively parallel architectures by 
\begin{enumerate*}[label=(\roman*)]
    \item re-encoding linearized indices for fast decoding on GPUs 
    and
    \item blocking the tensor into smaller chunks that fit in limited device memory and require at most 64 bits for indices.
\end{enumerate*}
However, these additional stages typically consume less than $25\%$ of the overall format construction cost, as shown in Figure~\ref{fig:breakdown}.

\vspace*{-2pt}
\section{Related Work}
\label{sec:related}
Optimizing sparse tensor decomposition and MTTKRP operations has been the subject of several prior studies, which propose various sparse tensor formats along with parallel algorithms to process and analyze the multi-modal data on CPU- and GPU-based hardware architectures. 
List-based formats, such as F-COO~\cite{liu2017unified}, GenTen~\cite{phipps2019software}, and TB-COO~\cite{Dun2021}, explicitly store the multi-dimensional coordinates of each non-zero element. 
To reduce atomic operations, these formats keep multiple mode-specific copies of the tensor and/or extra scheduling information, which substantially increases their memory footprint.
Tree-based formats, including  CSF~\cite{Smith2015, Smith2015a}, B-CSF~\cite{nisa2019load}, and MM-CSF~\cite{Nisa2019b}, extend the compressed sparse row (CSR) matrix format to higher-order tensors. 
While these mode-specific formats can compress the sparse data, they have load imbalance issues and significant performance variation across various modes of execution. 

Block-based formats (e.g., HiCOO~\cite{li2018hicoo}) cluster non-zero elements to compress the sparse tensor and to exploit data locality. However, as the number of tensor modes and sparsity increase, the majority of HiCOO blocks consist of a few non-zero elements, leading to more memory usage than COO~\cite{li2018hicoo, ahmed2021}. 
In addition, the irregular spatial distributions of sparse data result in severe load imbalance and synchronization issues across HiCOO blocks~\cite{li2018hicoo, ahmed2021}, and as a result, it has no GPU implementation~\cite{li_parti}.
In contrast, BLCO addresses these limitations by generating coarse-grained blocks that fit in the GPU memory, while efficiently utilizing the GPU resources by encoding the non-zero elements within each block in a fine-grained linearized form, amenable to caching and parallel execution.

CPU-based tensor linearization approaches, such as the LCO~\cite{harrison2018high} and ALTO~\cite{ahmed2021} formats, compress the tensor by mapping the multi-dimensional coordinates of a non-zero element into a single index.
In particular, ALTO enables high-performance tensor operations by 
\begin{enumerate*}[label=(\roman*)]
    \item leveraging the efficient bit-level scatter/gather instructions and large integers on CPUs, and
    \item using adaptive atomic- and reduction-based conflict resolution algorithms, tailored for architectures with coarse-grained cores/threads.
\end{enumerate*}   
However, massively parallel GPUs lack native support for the bit manipulation instructions and large integer arithmetic that are needed for efficient processing of prior linearized formats. Additionally, GPUs suffer from limited device memory and require sophisticated conflict resolution, due to their massive fine-grained parallelism and substantial synchronization overhead. 
BLCO directly addresses these issues to allow efficient execution of large-scale tensors on accelerators.  

Segmented scan and reduction~\cite{sengupta2008efficient,  yan2013streamscan} have been used to reduce the synchronization cost of sparse workloads~\cite{bell2009implementing, zhang2010fast, yan2014yaspmv, Liu2017, Dun2021} on parallel architectures.
Prior studies apply these  primitives to \emph{mode-specific} formats with delineated and/or sorted groups of non-zero elements according to the target mode. 
In contrast, we devise an opportunistic algorithm that leverages the \emph{mode-agnostic} BLCO format to reduce synchronization by discovering conflicts and performing segmented scan on-the-fly and without needing sorted non-zero elements or keeping extra scheduling information. 
In addition, our novel conflict resolution algorithm eliminates control-flow and memory-access irregularities by specializing threads to perform different operations at each execution phase.

The TACO compiler~\cite{kjolstad2017tensor} automatically generates various sparse matrix and tensor algebra kernels, including sparse MTTKRP. However, prior work showed that hand-optimized implementations of CSF-based formats (namely, B-CSF) still outperform the auto-generated TACO code on GPU architectures, even with extensive auto-scheduling and optimization~\cite{senanayake2020sparse}.

A wealth of work perform MTTKRP computations on distributed-memory platforms using MPI~\cite{7516087, 7832851, solomonik2013cyclops, Choi2014, shin2014distributed}, 
or the MapReduce~\cite{kang2012gigatensor, blanco_cstf_2018} framework. Other studies~\cite{zhao2018bridging, xie2019ia, sun2020sptfs} explore format selection based on machine learning models to efficiently leverage existing sparse formats.

\section{Conclusion and Future Work}
To enable high-performance sparse tensor decomposition on massively parallel GPU architectures, this work proposes the BLCO format. 
In contrast to prior approaches, which have been restricted to in-memory tensors, BLCO allows efficient processing of both in-memory and out-of-memory tensors. 
By discovering and merging conflicting updates on-the-fly, without any mode-specific information or ordering of non-zero elements, our BLCO-based MTTKRP demonstrated substantial speedup (up to $33.35\times$) over the state-of-the-art MM-CSF.  
Our future work will explore heterogeneous distributed-memory systems as well as other tensor algorithms.
\balance

\begin{acks}
The authors would like to thank Intel colleagues, John Pennycook and Nitin Dhamankar, for their help and valuable feedback.
\end{acks}

\bibliographystyle{ACM-Reference-Format}
\bibliography{header}

\end{document}